\pdfoutput=1
\documentclass[aps,prx,twocolumn,superscriptaddress,longbibliography,nofootinbib,10pt]{revtex4-2}
\usepackage{amsmath,amssymb,graphicx,siunitx,xcolor,hyperref}
\usepackage[T1]{fontenc}
\hypersetup{hidelinks}

\extrafloats{200}
\setcitestyle{super,open={},close={}} %
\newcommand{\eps}{\varepsilon}
\newcommand{\phx}{\texttt{ibm\_phoenix}}
\newcommand{\ins}[1]{\texttt{#1}}

\newcommand{\sfig}[1]{Fig.~\ref{#1}}
\newcommand{\snote}[1]{\rev{\hyperref[#1]{Appendix~\ref*{#1}}}}   %
\newcommand{\methref}[1]{\rev{\hyperref[#1]{Methods~\ref*{#1}}}}   %
\newcommand{\secref}[1]{\rev{\hyperref[#1]{Section~\ref*{#1}}}}    %
\makeatletter   %
\let\NCsubsection\subsection
\renewcommand{\subsection}[1]{\NCsubsection{#1}\NR@gettitle{#1}}
\makeatother
\newcommand{\mfig}[1]{Fig.~\ref{#1}}
\newcommand{\pan}[1]{(#1)}
\newcommand{\newsection}[1]{\section*{#1}\phantomsection\setcounter{subsection}{0}}   %
\newcommand{\rev}[1]{#1}   %
\newcommand{\revcolor}{}   %

\begin{document}
\title{Demonstration of Logical Error Suppression in a Surface Code Memory on a Cloud-Accessible IBM Superconducting Processor}

\author{Shuwen Kan}
\affiliation{Physical and Computational Sciences Division, Pacific Northwest National Laboratory}
\affiliation{Department of Computer and Information Science, Fordham University}

\author{Aaron Hoyt}
\affiliation{Physical and Computational Sciences Division, Pacific Northwest National Laboratory}
\affiliation{Department of Physics, University of Washington}

\author{Sean Garner}
\affiliation{Physical and Computational Sciences Division, Pacific Northwest National Laboratory}
\affiliation{Department of Physics, University of Washington}

\author{Ethan Vo}
\affiliation{Physical and Computational Sciences Division, Pacific Northwest National Laboratory}

\author{Adrian Harkness}
\affiliation{Physical and Computational Sciences Division, Pacific Northwest National Laboratory}

\author{Rod Rofougaran}
\affiliation{Physical and Computational Sciences Division, Pacific Northwest National Laboratory}

\author{Zefan Du}
\affiliation{Department of Computer and Information Science, Fordham University}

\author{Chenxu Liu}
\affiliation{Physical and Computational Sciences Division, Pacific Northwest National Laboratory}

\author{Norm Tubman}
\affiliation{NASA Ames Research Center}

\author{Ying Mao}
\affiliation{Department of Computer and Information Science, Fordham University}

\author{Samuel Stein}
\affiliation{Physical and Computational Sciences Division, Pacific Northwest National Laboratory}

\date{\today}

\begin{abstract}
Quantum error correction (QEC) can suppress logical errors exponentially with code distance when physical noise is below
threshold. Realizing this on hardware requires tight orchestration of a quantum processor, taxing gate
errors, mid-circuit measurements and qubit coherence. On real devices, qubit-to-qubit variation, noise asymmetry and
correlations across space and time are a small subset of challenges that complicate how these capabilities translate into logical error suppression. Here we study the performance of the rotated surface-code memory on an IBM Nighthawk r2 processor. We find heterogeneous performance across the processor when testing various placements which we attribute to varying qubit quality. We further optimize the performance of the code by combining short readout, fused measurement and reset, and an XX echo on data qubits during ancilla qubit readout. Relative to standard readout without reset or echo, the optimized syndrome extraction circuit lowers the per-round logical error rate ($\epsilon$) by median factors
of 3.8 and 1.4 in X and Z logical basis respectively. In
longer memory experiments of up to 150 rounds, the distance advantage weakens, alongside long-
lived correlations that may reflect the buildup of a persistent error component. These observations
connect circuit compilation and optimization design to the performance and current limitations of surface-code memories. Our observations align with motivating quantum processor and QEC circuit codesign for the execution of fault-tolerant quantum computing primitives as well as practical limitations of surface code memories on current hardware devices on the cloud. Under our optimization recipe, increasing the protected distance from 3 to 6 at fixed other distance reduces $\epsilon$ from 0.53 \% to 0.24 \% in Z memory and from 0.52 \% to 0.22 \% in X memory, an error-suppression factor ($\Lambda$) of 1.68 and 1.85 under maximum-likelihood decoding.

\end{abstract}

\maketitle

\newsection{Introduction}\label{sec:intro}

Quantum computers can solve problems that are intractable for classical computers
\cite{shor1997polynomial,reiher2017elucidating}. Realizing these applications, however, requires quantum error
correction (QEC), in which multiple physical qubits encode logical qubits to enable the detection and correction of certain errors. Below the relevant noise threshold, increasing the code distance can suppress
logical errors exponentially \cite{aharonov1997fault,knill1998resilient,dennis2002topological}. The surface code is a promising candidate for QEC on superconducting hardware due to its relatively high error threshold, efficient decoders and requirement for only nearest-neighbour interactions on a two-dimensional square lattice \cite{kitaev2003fault,bravyi1998quantum,raussendorf2007fault,fowler2012surface,higgott2025sparse}. Experiments have demonstrated repeated syndrome extraction \cite{krinner2022realizing,zhao2022realization}, improved logical performance with increasing distance \cite{google2023suppressing}, and operation below the surface-code threshold \cite{google2025below}. Surface-code encoding and logical operations have also been demonstrated with trapped ions \cite{erhard2021entangling} and reconfigurable neutral-atom arrays \cite{bluvstein2024logical}. Related demonstrations include four-dimensional surface codes in trapped ions \cite{berthusen2024experiments}, fault-tolerant operation of other trapped-ion codes \cite{egan2020fault,ryan2021realization} and a fault-tolerant neutral-atom architecture \cite{bluvstein2026fault}.

\begin{figure*}[!tb]
\centering
\includegraphics[width=\linewidth]{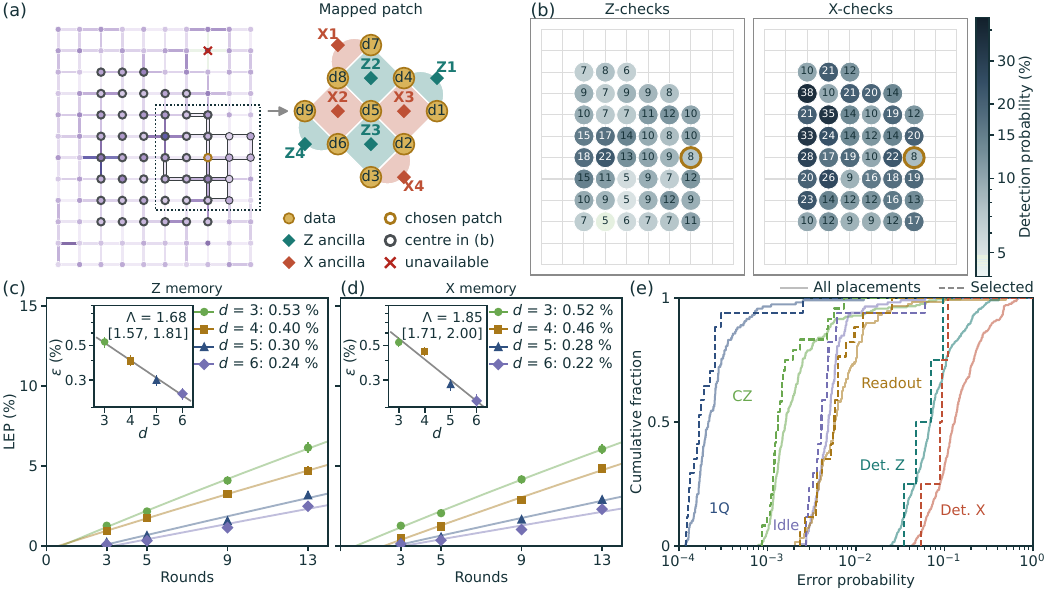}
\caption{\textbf{Device and rotated surface-code performance.}
Increasing the protected distance lowers the logical error per round in both memories for the selected rectangular patches.
\rev{\pan{a} The processor and selected placement, with qubits coloured by readout error and couplers by controlled-Z (CZ) gate error.}
\pan{b} Check detection probabilities for 44 placements, pooling both memories.
\rev{\pan{c},~\pan{d} Z- and X-memory logical error probability (LEP) versus rounds at other distance 3, using exact maximum-likelihood decoding on all \num{20000} shots.}
Insets show $\eps$ versus protected distance, with fits used to estimate the suppression factor $\Lambda$.
\rev{\pan{e} Calibration and detection-probability distributions, with dashed curves for the selected placement.}
Detection distributions use Z-memory records, and the idle-dephasing proxy uses a \SI{1.512}{\micro\second} interval.
\rev{Error bars on LEP are two-sided 95\,\% Wilson intervals. Inset $\eps$ and $\Lambda$ intervals use 400 binomial resamples with the placements fixed (\snote{sn:stats}).}}
\label{fig:device}
\end{figure*}

The implementation of quantum error correction taxes all aspects of a quantum processor's capabilities, including but not limited to single- and two-qubit gate performance, mid-circuit measurements and qubit coherence. Variation across qubits, noise asymmetry and correlations across space
and time complicate how these capabilities translate into the logical error suppression attainable. On cloud-accessible processors, the available optimizations are further shaped by the provider’s native instructions, device information, and scheduling tools. On the IBM Nighthawk r2 processor studied here,
standard ancilla readout occupies approximately 80\,\% of the baseline syndrome-extraction round, taxing the data qubits coherence during this interval.

Previous experiments have shown how circuit choices materially affect error-correction performance. Early cloud experiments ran single-round repetition codes \cite{wootton2018repetition} and a post-selected $[[4,2,2]]$ code \cite{harper2019fault}. Prior `heavy-hex' layouts with weight-3 connectivity cannot host the surface code's weight-four checks natively \cite{chamberland2020topological}, thus prior experiments used QEC codes adapted to its connectivity \cite{chen2022calibrated,sundaresan2023demonstrating,hetenyi2024creating}. On IBM heavy-hex processors, Vezvaee et al. combined a SWAP-based surface-code embedding with dynamical decoupling and showed that the interpretation of apparent distance scaling can depend on the decoupling applied to the compared codes \cite{vezvaee2026surface}. Harper et al. found readout idling to be highly detrimental to a heavy-hex memory and improved its performance by shortening the syndrome-extraction circuit \cite{harper2026characterising}. Surface-code circuits on IQM's Garnet allocated fresh ancillas in every round \cite{Krzyszkowski_2025}. Experiments on native square-lattice IBM processors have also examined logical-error estimation from syndrome data and surface-code operation with defective components \cite{takou2026logical,drouet2026defective,kim2026luci}. Native connectivity reduces routing requirements, while mid-circuit reset expands the choices for ancilla reuse. Together, these capabilities allow readout, reset and idle scheduling to be compared within the same memory experiment.

Here we study rotated surface-code memories on the IBM Nighthawk r2 processor \texttt{ibm\_phoenix} (\mfig{fig:device}), which provides native square connectivity and mid-circuit reset \cite{ibm2026nighthawkr2}. With exact maximum-likelihood decoding on a calibration-derived noise model, built for each executed circuit from that day's device calibration and the executed schedule rather than fitted to the decoded shots (\methref{sec:m_noise}), increasing the protected distance of rectangular patches from 3 to 6 reduces the logical error per round $\varepsilon$ by factors of approximately 2.2 in Z memory and 2.4 in X memory. The dual operator distance is fixed at 3, and $\varepsilon$ is fitted over 3-13 rounds. Comparisons of readout durations, reset strategies and dynamical decoupling sequences guide a jointly optimized circuit recipe combining short readout, fused measurement and reset, and an echo on waiting data qubits. Under minimum-weight perfect matching, the recipe lowers $\varepsilon$ by median factors of 3.8 in X memory and 1.4 in Z memory relative to standard readout without reset or echo on the same qubits. Every individual 95\,\% gain interval lies above unity (\mfig{fig:recipe_gain}). The larger improvement in X memory is consistent with the reductions in phase-error signatures observed when the readout and idle controls are changed.

The distance advantage weakens in longer memories. In separate experiments extending to 150 rounds, the logical error per round increases alongside long-lived syndrome correlations. The syndrome records show a buildup of persistent detector correlations during the memory, hinting at cumulative non-Markovian error channels. Our results connect changes to syndrome extraction with their logical benefit and examine how that benefit changes as the code distance increases and the memory is extended in time.

\begin{figure*}[!tb]
\centering
\includegraphics[width=\linewidth]{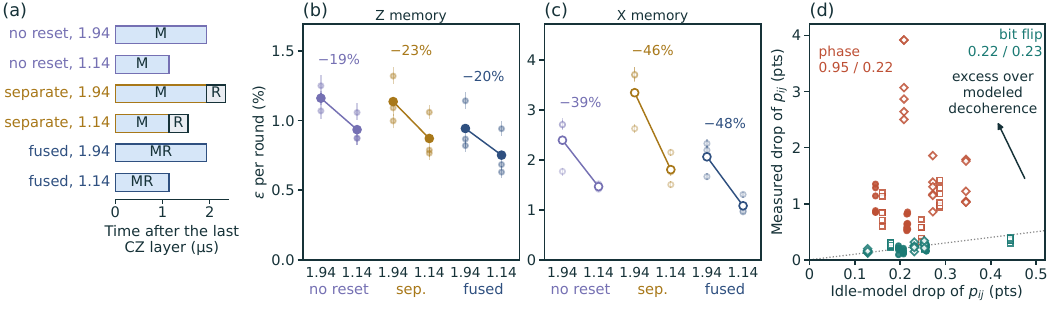}
\caption{\textbf{Readout-window duration.}
Shorter readout lowers logical errors and space-like pair probabilities, with phase-pair changes exceeding the idle-decoherence prediction.
\pan{a} Ancilla schedules with no reset, separate reset or fused measurement and reset.
\pan{b},~\pan{c} Z- and X-memory $\eps$ versus readout duration.
\rev{Faint points show individual placements and prominent markers their means.} Labels give fractional changes.
\pan{d} Measured reductions in space-like $p_{ij}$ versus predictions from the shared data qubit's calibrated $T_1$ or $T_2$.
Markers distinguish placements, the dotted line denotes equality, and annotations give measured and predicted medians.
\rev{Error bars in (b,c) are 95\,\% intervals from 200 binomial resamples of each round count, with the placements' half-widths combined in quadrature for the means (\snote{sn:stats}).}}
\label{fig:window}
\end{figure*}

\newsection{Results}\label{sec:results}

\subsection{Surface-code memories and hardware heterogeneity}\label{sec:memories}

We implement rotated surface-code memories without SWAP gates on the native square connectivity of the IBM Nighthawk r2 processor \phx{}, shown in \mfig{fig:device}a. Each syndrome-extraction round comprises \rev{four layers of controlled-Z (CZ) gates} followed by ancilla readout and, where used, reset (\methref{sec:m_device}). The Z- and X-memory experiments begin in the logical states \(|0\rangle\) and \(|+\rangle\), respectively. For rectangular patches, we distinguish the protected distance of the stored memory basis from the other distance. We report logical error probability (LEP) after a specified number of rounds and logical error per round \(\varepsilon\), fitted over 3, 5, 9 and 13 rounds (\methref{sec:m_noise}).

Memory performance varies strongly with placement. The detection maps in \mfig{fig:device}b show spatial variation across the processor. Across 88 usable distance-three placements, five-round LEP ranges from 3.4\,\% to 17.2\,\% in Z memory and from 6.0\,\% to 39.8\,\% in X memory, using standard readout without reset or echo. The detection probability of the relevant check family correlates with LEP, with Spearman coefficients of 0.86 for Z memory and 0.89 for X memory (\sfig{fig:scan_det_lep}a,b; \snote{sn:scan}). We use the scan ranking to choose a characterization placement and two further placements that share no qubits with it or with each other.

Increasing the protected distance improves short-memory performance on the selected rectangular patches. With the dual distance fixed at 3, increasing the protected distance from 3 to 6 lowers \(\varepsilon_Z\) from 0.526\,\% [0.481\,\%, 0.565\,\%] to 0.244\,\% [0.224\,\%, 0.266\,\%], and \(\varepsilon_X\) from 0.523\,\% [0.486\,\%, 0.564\,\%] to 0.220\,\% [0.201\,\%, 0.244\,\%] (\mfig{fig:device}c,d). These measurements use the optimized circuit described in \methref{sec:m_protocols}, combining \ins{measure\_reset\_2} with a data-qubit echo. \rev{Placements are selected using calibration-derived minimum-weight perfect matching (MWPM), whereas the displayed curves use exact maximum-likelihood decoding on all 20,000 shots per circuit.}

\rev{Rotating the circuit gives similar performance when the same physical direction is protected.} Alongside the
component variation shown in \mfig{fig:device}e, we compare placements using the same qubits with exchanged X-
and Z-check roles (\sfig{sfig:transpose}a). \rev{Across 60 optimized-circuit comparisons spanning the tested distance pairs, MWPM gives a median error ratio of 1.01, with an interquartile range of 0.94--1.05, and 48 of the 60 ratios have intervals containing unity (\sfig{sfig:transpose}b; \snote{sn:transpose}).} Thus, typical performance changes little under this
exchange of memory basis and check roles.

\subsection{Readout duration}\label{sec:readout}

In this section and \rev{in} \secref{sec:reset}, we compare readout durations and reset schemes on the three distance-three placements
chosen from the placement scan. Ancilla readout leaves the data qubits idling for the majority of each syndrome extraction round. Standard readout occupies \SI{1.94}{\micro\second} of an approximately \SI{2.39}{\micro\second} round, or about 80\,\% of its duration. \mfig{fig:window}a shows the schedules we compare: standard and \SI{1.14}{\micro\second} readout, each without reset, with separate reset or with fused measurement and reset. Shorter readout lowers every point estimate of \(\varepsilon\) across the 18 placement, memory and reset comparisons. On the selected placement with fused reset, \(\varepsilon_Z\) falls from 0.82\,\% to 0.63\,\%, a relative reduction of 23\,\%. The corresponding \(\varepsilon_X\) falls from 1.67\,\% to 0.98\,\%, a reduction of 41\,\% (\mfig{fig:window}b,c; \snote{sn:readout_det}).

\begin{figure*}[!tb]
\centering
\includegraphics[width=\linewidth]{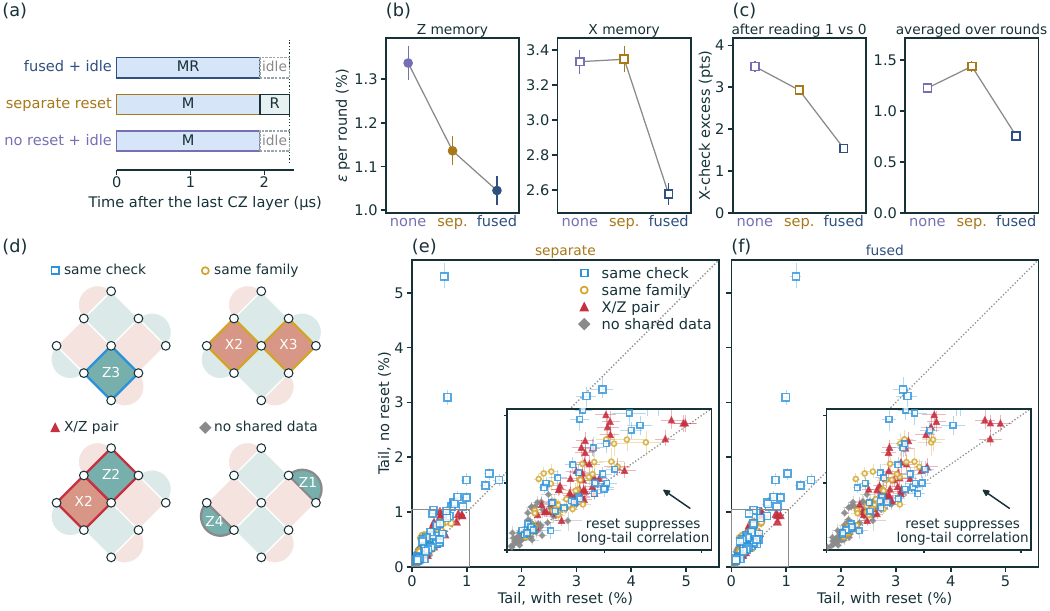}
\caption{\textbf{Reset at equal round duration.}
Both reset schemes reduce long-tail correlations, while fused reset also lowers X-memory logical error at matched round duration.
\pan{a} Ancilla schedules padded to a common \SI{2.79}{\micro\second} round.
\pan{b} Mean $\eps$ across three placements.
\pan{c} Excess neighbouring X-check activity after recorded ancilla readout 1 versus 0, relative to non-neighbouring checks, before and after weighting by the read-1 fraction.
\pan{d} Detector-pair classes.
\pan{e},~\pan{f} Long-tail $p_{ij}$ without reset versus separate and fused reset, respectively.
\rev{Points above equality have smaller correlations with reset. Insets enlarge the low-correlation region.}
Tail offsets account for each circuit's single-fault reach.
Error bars are 95\,\% intervals: in (b) from binomial resampling, combined in quadrature over placements (\snote{sn:stats}); in (c) from binomial counting of detector samples (\snote{sn:conditioned}); in (e,f) from 200 whole-shot bootstrap resamples.}
\label{fig:zz}
\end{figure*}

To identify which error sources limit these circuits, we form an error budget \cite{google2025below}: in simulations, each group of noise sources is scaled in turn, and its
effect on \(\varepsilon\) gives its contribution (\snote{sn:reset_budget}). The error budget assigns its largest contribution to
data-qubit idling during readout (\sfig{sfig:reset_budget}). In the long-readout circuit without reset, this accounts for 57\,\% and 63\,\% of modelled \(\varepsilon\) in Z and X memory respectively, and decreases with the short fused-readout window (\sfig{sfig:reset_budget}a,b). \rev{The budget captures part of the measured improvement (\snote{sn:reset_budget}).}

As shown in \mfig{fig:window}d, idle decoherence predicted from the reported \(T_1\) and \(T_2\) of the data qubits accounts for the bit-flip reduction but under predicts the phase-associated reduction. We infer effective pair probabilities \(p_{ij}\) from detector correlations \cite{google2021exponential}. Space-like pairs connect checks sharing a data qubit: Z-check pairs are sensitive to bit flips, and X-check pairs to phase errors (\snote{sn:dem}). The prediction treats the \SI{0.80}{\micro\second} shorter wait as Pauli-twirled relaxation and dephasing of the data qubit shared by each space-like detector pair (\snote{sn:idle_pred}). For this shorter wait, the median measured decrease in bit-flip-associated \(p_{ij}\) is 0.22 percentage points, close to the 0.23 points predicted from \(T_1\). The phase-associated decrease is 0.95 points, compared with 0.22 points predicted from \(T_2\) (\mfig{fig:window}d). This additional phase sensitivity is consistent with idle-time effects reported on IBM heavy-hex processors \cite{vezvaee2026surface}. \rev{Whereas residual-ZZ dephasing of spectator qubits during measurement has been described before\cite{hashim2025efficient,govia2023randomized}, here we compare it across readout and reset instructions and test it in the syndrome records.} In a separate spectator control experiment, in which one ancilla was held in \(|0\rangle\) or \(|1\rangle\) while we tracked the phase of its neighbouring data qubits over time, these data qubits accumulate phase at 0.119 \(\mathrm{rad}\,\mu\mathrm{s}^{-1}\) with the ancilla in \(|0\rangle\) and 0.199 \(\mathrm{rad}\,\mu\mathrm{s}^{-1}\) with it in \(|1\rangle\). The difference, 0.079 \(\mathrm{rad}\,\mu\mathrm{s}^{-1}\), is consistent with a contribution from residual ZZ coupling (\sfig{sfig:spectator}a,b). The phase a data qubit gains during readout therefore depends on the state of its neighbouring ancilla.

\begin{figure*}[!tb]
\centering
\includegraphics[width=\linewidth]{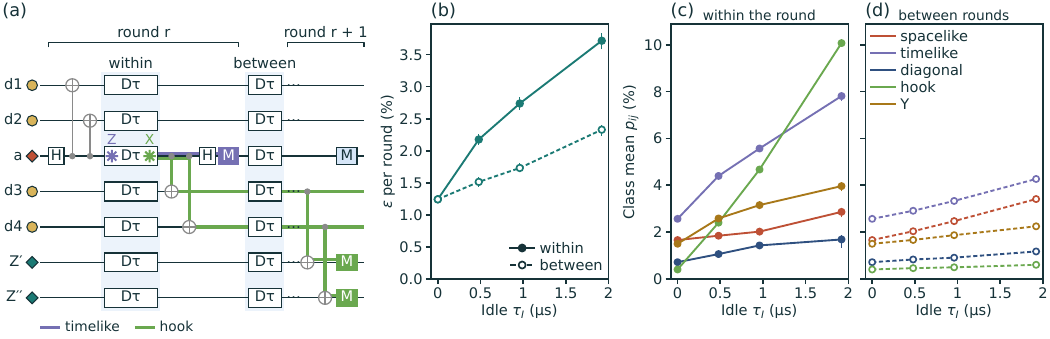}
\caption{\textbf{Idle placement within syndrome extraction.}
The same inserted delay produces larger logical errors and hook-class correlations within the round than between rounds.
\pan{a} Delay positions after the second entangling layer or after ancilla readout, with illustrative fault propagation.
\pan{b} Z-memory $\eps$ versus inserted delay: solid lines and filled markers denote within-round insertion; dashed lines and open markers denote between-round insertion.
\pan{c},~\pan{d} Corresponding class-averaged $p_{ij}$, using a common scale.
\rev{Measurements use the selected placement. \sfig{sfig:idle_x} shows the complementary X-memory comparison.}
Error bars are 95\,\% intervals from 200 binomial resamples of each round count (b) and from 200 whole-shot bootstrap resamples (c,d).}
\label{fig:idle}
\end{figure*}

\subsection{Ancilla reset at equal round duration}\label{sec:reset}

Separate reset lengthens the round by \SI{0.40}{\micro\second}, so comparing the reset schemes at their natural
durations would mix the effect of reset with that of a longer idle. To isolate the reset strategy, we pad the no-reset and fused-reset schedules to the same approximately \SI{2.79}{\micro\second} round as separate reset, giving the three schedules in \mfig{fig:zz}a. Fused reset performs best: while both reset schemes reduce persistent detector correlations, fused
reset also most shortens the time an ancilla spends in \(|1\rangle\) after readout. Across three placements, fused reset
lowers the mean \(\varepsilon_Z\) from 1.34\,\% to 1.05\,\%, and the mean \(\varepsilon_X\) from 3.33\,\% to 2.58\,\% (\mfig{fig:zz}b). Separate reset helps only Z memory: it lowers the mean \(\varepsilon_Z\) to 1.14\,\%,
whereas the mean \(\varepsilon_X\) stays at 3.35\,\%, indistinguishable from the 3.33\,\% without
reset.

The spectator experiment is consistent with an explanation for this difference. An ancilla in \(|1\rangle\) adds
phase to its neighbouring data qubits (\sfig{sfig:spectator}), and phase errors limit X memory. Separate reset
shortens the inferred interval the ancilla spends in \(|1\rangle\) (2.11 versus \SI{2.40}{\micro\second} without
reset) but yields a read-1 outcome in more rounds (49\,\% versus 35\,\%). In a simple ZZ model the phase error grows
with the square of the interval, so these changes largely cancel. Fused reset shortens the interval to
\SI{1.48}{\micro\second} and halves the predicted phase contribution (\snote{sn:exposure}).

\mfig{fig:zz}c probes this explanation in the syndrome records. If an ancilla in \(|1\rangle\) adds phase
to its data neighbours, the X checks sharing those data qubits should fire more often in the round after the ancilla
reads 1 than after it reads 0. For the bulk Z-check ancillas in the X-memory records, we compare these next-round firing
probabilities and subtract the same comparison for X checks sharing no data qubits with the ancilla. Weighted
by how often the ancilla reads 1, this net X-check excess is 1.23 percentage points without reset, 1.44 with separate
reset and 0.75 with fused reset (\snote{sn:conditioned}), the same ordering as the predicted phase contribution: highest
with separate reset and lowest with fused reset. The simple ZZ estimate reproduces this ordering but underpredicts the
absolute excesses by approximately a factor of two (\snote{sn:exposure}).

Both reset schemes also provide evidence of reduced persistent correlations, quantified by long-tail estimates of \(p_{ij}\) (\methref{sec:m_correlations}). These correlations extend across several rounds, beyond the temporal reach of a single fault in the modeled circuit (\sfig{sfig:tail3d}). \mfig{fig:zz}d--f distinguish same-check pairs, same-family pairs sharing a data qubit, X/Z-check pairs sharing data qubits, and pairs sharing no data qubits. \rev{The grey points represent pairs sharing no data qubits, whose long-tail correlations may reflect broader noise contributions, including shot-to-shot fluctuations.} Some large same-check tails decrease strongly (\snote{sn:reset_tail}). \rev{Round-by-round correlation maps for individual check pairs show how the persistent correlations and their response to reset vary across rounds and pair types (\snote{sn:pairmaps}; \sfig{sfig:pairmaps}b--e).}

\begin{figure*}[!tb]
\centering
\includegraphics[width=\linewidth]{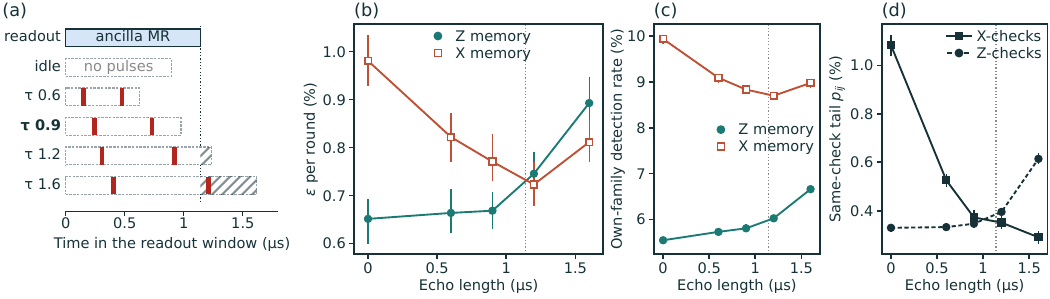}
\caption{\textbf{Dynamical decoupling during readout.}
The echo lowers X-memory logical error, while extending it beyond readout raises the error in both memories.
\rev{\pan{a} X--X sequences during fused readout, with hatched segments beyond the readout window.}
\rev{\pan{b} $\eps$ versus nominal echo length in Z and X memory (zero: no-echo control).}
\pan{c} Detection probability of each memory's check family.
\pan{d} Same-check long-tail $p_{ij}$, averaged over both memories, for X-checks and Z-checks.
Dotted lines mark the readout duration.
\rev{Nominal lengths exclude pulse durations. Longer sequences extend the enclosing round.}
Error bars are 95\,\% intervals from 200 binomial resamples of each round count (b) \rev{or from 200 whole-shot bootstrap resamples (c,d). In (d), the two memories' half-widths are combined in quadrature (\snote{sn:stats}).}}
\label{fig:echo}
\end{figure*}

\subsection{Idle errors and dynamical decoupling}\label{sec:idle}

The readout results motivate examining whether the cost of idle time also depends on its position in the round. \mfig{fig:idle}a shows the two positions of a pulse-free delay inserted on every qubit: after the second entangling layer or after ancilla readout. For a \SI{1.92}{\micro\second} delay, the Z-memory error per round, \(\varepsilon\), rises
from 1.24\,\% without the added delay to 3.72\,\% within the round,
compared with 2.33\,\% between rounds (\mfig{fig:idle}b). The X memory shows the same
ordering (\sfig{sfig:idle_x}a,b). Thus, the same added waiting time is more costly when it interrupts syndrome
extraction.

The larger within-round penalty accompanies a strong increase in hook-class correlations. In the
executed-circuit model, a hook fault occurs when an ancilla fault propagates through later entangling gates onto
data qubits. In Z memory, the mean \(p_{ij}\) of the associated detector pairs reaches 10.07\,\% with the within-round delay, compared with 0.61\,\% for the same delay between
rounds (\mfig{fig:idle}c,d). X memory shows a similar response (\sfig{sfig:idle_x}c,d). This enhancement is
consistent with the hook-error patterns of the modeled circuit.

The phase-sensitive errors observed during readout motivate utilizing dynamical decoupling over the idling data qubits. Dynamical decoupling has been used for idle error suppression in surface-code experiments on Google and IBM processors
\cite{google2023suppressing,google2025below,vezvaee2026surface}. \mfig{fig:echo}a shows the X--X sequence we apply during the short fused-readout window. On the selected placement, the nominal \SI{0.9}{\micro\second} echo
lowers \(\varepsilon_X\) from 0.981\,\% to 0.771\,\%, a relative
reduction of 21\,\%. The Z-memory point estimate changes from 0.651\,\% to 0.668\,\%,
an increase of 2.6\,\% with overlapping intervals (\mfig{fig:echo}b). The X-memory
benefit is reproduced on a second placement, where the reduction is 38\,\% (\sfig{sfig:echo_p35}a,b).

\begin{figure}[!tb]
\centering
\includegraphics[width=0.7\linewidth]{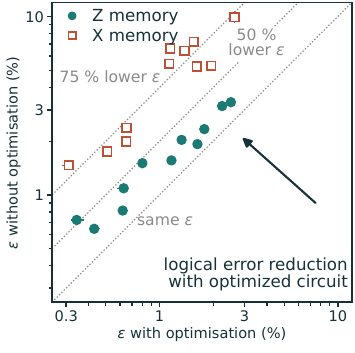}
\caption{\textbf{Recipe gains across patch shapes.}
The recipe lowers $\eps$ for all eleven tested distance pairs in both memories.
Each point compares the same qubits, with recipe $\eps$ on the horizontal axis and baseline $\eps$ on the vertical axis. Filled circles denote Z memory and open squares X memory. The baseline uses standard readout without reset or echo. \rev{Both axes are logarithmic. Dotted guides denote equal $\eps$ and reductions of 50\,\% and 75\,\%.} The arrow indicates increasing benefit from the recipe.
Error bars are 95\,\% intervals from 200 binomial resamples of each round count.}
\label{fig:recipe_gain}
\label{sfig:recipe_gain}
\end{figure}

\begin{figure*}[!tb]
\centering
\includegraphics[width=\linewidth]{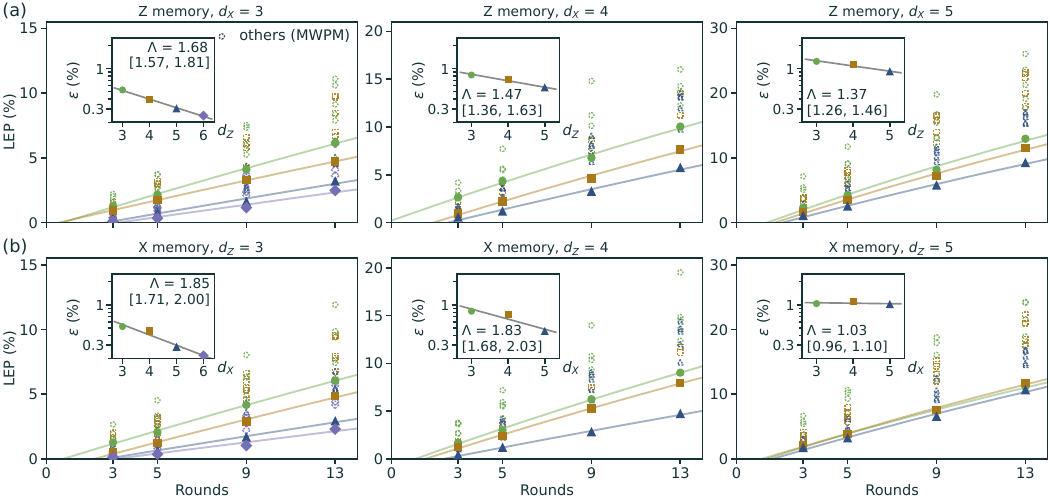}
\caption{\textbf{Memory performance across patch geometries.}
Logical error decreases with protected distance in both memories at other distances 3 and 4, and in Z memory at
other distance 5.
\pan{a},~\pan{b} Z- and X-memory LEP versus rounds, with columns fixing the other distance at 3, 4 or 5.
Prominent curves show the placement with the lowest $\eps$ for each patch shape.
These curves use exact maximum-likelihood decoding on all \num{20000} shots.
Faint points show other placements decoded by MWPM.
Insets show $\eps$ and fitted $\Lambda$.
\rev{Error bars on LEP, omitted for faint points, are two-sided 95\,\% Wilson intervals.} Inset $\eps$ and $\Lambda$ intervals use 400 binomial resamples with the placements fixed (\snote{sn:stats}).}
\label{fig:scaling}
\end{figure*}

Extending the echo eventually increases errors in both memories. On the selected placement, increasing the nominal duration from 1.2 to \SI{1.6}{\micro\second} raises \(\varepsilon_Z\) by 20\,\% and \(\varepsilon_X\) by 12\,\% (\mfig{fig:echo}b). These longer sequences extend the round beyond the readout window and add waiting time for both data and ancilla qubits. The duration sweep therefore changes both the decoupling sequence and the total round duration.

Detector statistics show the accompanying trade-off. With the \SI{0.9}{\micro\second} echo, the Z-check detection probability rises from 5.54\,\% to 5.81\,\%, a relative increase of 4.7\,\%, while the X-check detection probability falls by 11\,\% (\mfig{fig:echo}c). The X-check long tail decreases from 1.08\,\% to 0.37\,\%, a reduction of approximately 66\,\% (\mfig{fig:echo}d). These observations are consistent with reduced phase-sensitive errors accompanied by additional bit-flip-sensitive detection events and a small change in the Z-memory \(\varepsilon\) that lies within its interval. In repeated readout windows outside the code, the echo lowers the median flip probability from \(|+\rangle\) after eight windows from 14.7\,\% to 4.1\,\%, while raising it from \(|0\rangle\) from 0.2\,\% to 3.5\,\% (\snote{sn:justdd}; \sfig{sfig:justdd}).
Together, the measurements motivate protecting the data qubits during the readout idle while limiting the additional circuit duration.

\subsection{Circuit optimization, scaling and post-selection}\label{sec:optimization}

We extend the measurements to eleven rotated surface-code patch shapes, including \((d_X,d_Z)=(3,6)\) and \((6,3)\). As \mfig{fig:recipe_gain} shows, combining short readout, fused reset and a data-qubit echo improves logical performance across these shapes. The optimized circuit uses fused short measurement and reset with a nominal \SI{1.1}{\micro\second} X--X echo that fits within the readout window. Relative to standard readout without reset or echo on the same qubits, it reduces \(\varepsilon\) by a median of 29\,\% [24\,\%, 42\,\%] in Z memory and 73\,\% [69\,\%, 79\,\%] in X memory. All 22 matched comparisons favour the optimized circuit, with
individual intervals excluding no improvement.

The other distance affects the absolute logical error as well as the benefit of increasing the protected distance. \rev{At protected distance 3, increasing the other distance from 3 to 5 raises \(\varepsilon_Z\) from 0.526\,\% to 1.238\,\%, and X memory changes in the same direction (\mfig{fig:scaling}a,b).} Widening the patch introduces additional physical qubits and gate operations without increasing the protected distance. It also changes the hardware components used, so this comparison does not separate the contributions of circuit size and local component quality.

Increasing the protected distance improves performance at several fixed other distances. At other distance 3, the fitted two-distance-step suppression factors are \(\Lambda=1.68\) [1.57, 1.81] in Z memory and 1.85 [1.71, 2.00] in X memory (\mfig{fig:scaling}a,b, left column). Both fitted trends also decrease at other distance 4. At other distance 5, the fitted factors are 1.37 [1.26, 1.46] in Z memory and 1.03 [0.96, 1.10] in X memory.

\begin{figure*}[!tb]
\centering
\includegraphics[width=\linewidth]{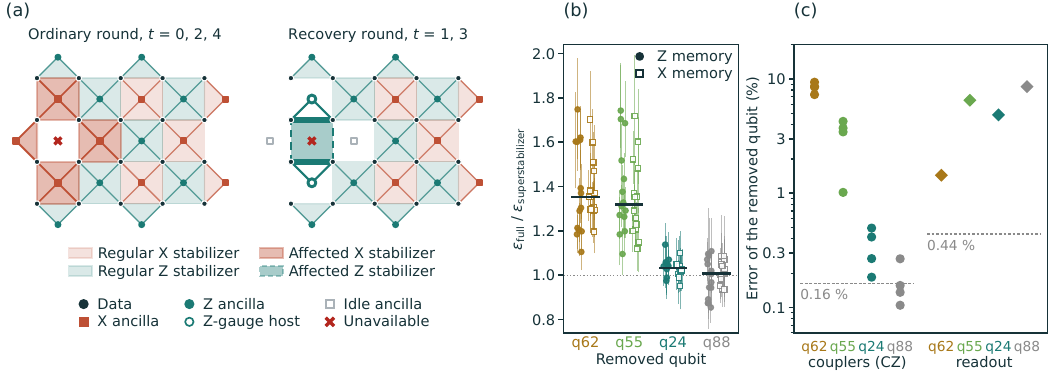}
\caption{\textbf{Superstabilizer layouts around weak qubits.}
\rev{Removing an ancilla with poor CZ couplers improves logical performance, whereas removing one with poor readout changes little.}
\rev{\pan{a} Ordinary and recovery rounds around a removed qubit. Alternating gauge measurements reconstruct stabilizer information.}
\pan{b} Full-patch $\eps$ divided by superstabilizer-layout $\eps$ on the same placement.
Filled and open markers denote Z and X memory, colours identify the removed qubit, and horizontal marks show medians.
\rev{\pan{c} Calibrated CZ and readout errors of the removed qubits, with dashed lines at the device medians.}
Both layouts use the recipe.
Error bars propagate the 95\,\% $\eps$ intervals of the two layouts to their ratio (\snote{sn:stats}).}
\label{fig:superstab}
\end{figure*}

\begin{figure*}[!tb]
\centering
\includegraphics[width=\linewidth]{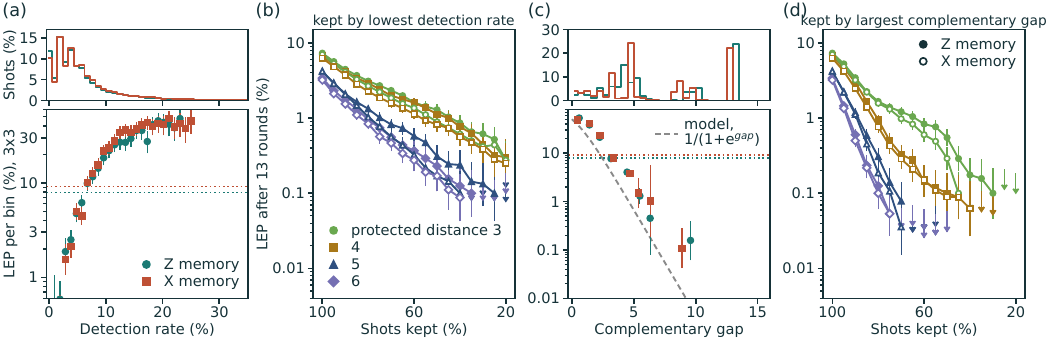}
\caption{\textbf{Detector statistics and post-selection.}
Larger complementary gaps identify shots with lower observed failure probability.
\pan{a} Shot fractions and conditional LEP after 13 rounds versus detection rate for a distance-three memory with a nominal \SI{0.9}{\micro\second} echo.
\pan{b} Conditional LEP after 13 rounds when retaining the shots with the lowest detection rate, for protected distances 3--6 with a \SI{1.1}{\micro\second} echo.
\rev{\pan{c} The shots of (a), binned by complementary gap, with $1/(1+e^g)$ dashed.}
\pan{d} The shots of (b), retaining the largest gaps.
\rev{Down-carets mark upper endpoints of two-sided 95\,\% Wilson intervals for one to four failures. Zero-failure points are omitted.}}
\label{fig:postsel}
\end{figure*}

Larger patches can include ancillas with elevated calibrated readout errors or higher CZ error rates, even when these components remain operational. We test whether excluding four such ancillas through
superstabilizer circuits improves performance \cite{leroux2025snakes,drouet2026defective}. In the layouts of \mfig{fig:superstab}a, neighbouring ancillas measure gauge operators on alternating rounds, and their outcomes are combined to recover the affected stabilizer information without using the excluded ancilla or its couplers (\snote{sn:superstab}). Removing q62 or q55 reduces
\(\varepsilon\) by a median of approximately 26\,\% and 24\,\%, respectively, with every individual interval
excluding zero reduction across 28 comparisons for each removal (\mfig{fig:superstab}b). These ancillas have poorer CZ gate performance than the other tested removals (\mfig{fig:superstab}c). Removing q24 or q88, which have poor readout but better CZ gate performance, gives median reductions of approximately 3\,\% and 1\,\% and little consistent benefit. These comparisons suggest considering coupler quality alongside ancilla readout when choosing placements.

In addition, we test post-selection with two scores, each meant to discard the shots most likely to be decoded
incorrectly. The first uses the raw detection statistics: the detection rate of the shot (\mfig{fig:postsel}a,b). The
second is the complementary gap \(g\), a measure of decoding confidence
\cite{bombin2024fault,gidney2025yoked,smith2024mitigating}. The MWPM decoder based on the calibration-derived detector
error model is constrained to each of the two logical outcomes in turn, and the gap value is the difference between the
weights of the two resulting predicted errors. \rev{The gap is small when both outcomes are nearly equally likely and large when one is much more likely (\methref{sec:m_noise}; \mfig{fig:postsel}c,d).}
\rev{\mfig{fig:postsel}a,c show how the shots are distributed over each score and the LEP within each bin. \mfig{fig:postsel}b,d show} the conditional LEP after retaining a given fraction of the shots, ranked by each score, for
protected distances 3--6. LEP increases with detection rate, and rate-based selection lowers conditional LEP after 13
rounds (\mfig{fig:postsel}a,b). Observed failure probability decreases with increasing gap, although it remains above
the matching-based estimate \(1/(1+e^g)\) over the displayed bins (\mfig{fig:postsel}c). The gap therefore provides
useful relative confidence despite its imperfect absolute calibration.

Gap-based selection gives lower conditional LEP than rate-based selection at the same retained fraction. At 60\,\% retention, the distance-three Z-memory LEP is 0.83\,\% with gap selection, \rev{compared with 1.62\,\% with rate selection. X memory shows the same ordering (\mfig{fig:postsel}d).} At protected distance 6, gap selection leaves one Z-memory failure and two X-memory failures among 12,000 retained shots. The upper endpoints of their two-sided Wilson intervals are 0.047\,\% and 0.061\,\%, respectively. These sparse counts limit the precision of the conditional estimates. The reduction is obtained by discarding 40\,\% of the shots and is distinct from the unconditional circuit and distance improvements reported above.

\begin{figure*}[!tb]
\centering
\includegraphics[width=\linewidth]{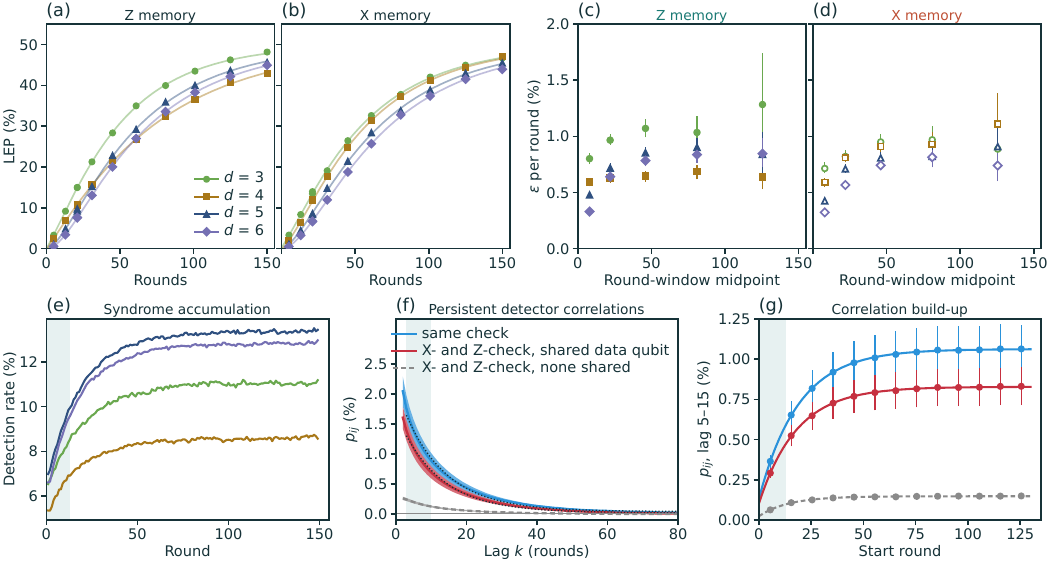}
\caption{\textbf{Long-memory performance and detector correlations.}
The distance advantage weakens in longer memories, alongside persistent detector correlations.
\pan{a},~\pan{b} Z- and X-memory LEP with time-dependent fits, for protected distances 3--6 at other distance 3.
\pan{c},~\pan{d} Corresponding $\eps$ estimates from successive round windows.
\pan{e} Detection probability averaged over both memories.
\pan{f} $p_{ij}$ versus lag for the same check and mixed-family pairs \rev{with or without shared data qubits, with exponential fits dotted.}
\pan{g} Correlations at lags 5--15 versus starting round, with fits to their buildup.
All measurements use recipe circuits.
Error bars are two-sided 95\,\% Wilson intervals in (a,b) and 95\,\% intervals from 200 binomial resamples of both endpoint round counts in (c,d), omitted where fewer than 97.5\,\% of resamples give a finite estimate. Bands and bars in (f,g) span the mean plus or minus 1.96 standard errors across 22 patch--memory combinations (\snote{sn:stats}).}
\label{fig:longmem}
\end{figure*}

\subsection{Long-memory performance}\label{sec:longmem}

Longer memories test whether the distance advantage observed over 3--13 rounds persists in time. \mfig{fig:longmem}a,b show the LEP of the optimized circuit for up to 150 rounds on selected placements with protected distances 3--6 and other distance 3, with all shots retained and decoded by calibration-derived MWPM. Within each memory, we compare logical error per round across successive round windows.

The largest tested protected distance shows a substantial increase in windowed \(\varepsilon\). Between rounds 3--13 and 61--101, the distance-six Z-memory estimate rises from 0.330\,\% [0.306\,\%, 0.358\,\%] to 0.838\,\% [0.767\,\%, 0.927\,\%], and the X-memory estimate from 0.323\,\% [0.299\,\%, 0.351\,\%] to 0.815\,\% [0.725\,\%, 0.882\,\%] (\mfig{fig:longmem}c,d). These are increases of approximately 2.5 in both memories. The corresponding distance-three estimates increase less, so the point estimate of the distance-three-to-six reduction factor falls from approximately 2.4 to 1.2 in Z memory and from 2.2 to 1.2 in X memory.

The weakening distance advantage accompanies changes in the raw syndrome records. Bulk-check detection probabilities rise during the memory (\mfig{fig:longmem}e), in nearly every check (\sfig{sfig:longcheck}), \rev{consistent with reports for surface-code memories without data-qubit leakage removal} \cite{google2025below}. Same-check correlations and mixed-family correlations between checks sharing a data qubit persist across many rounds, whereas pairs without shared data qubits are less strongly correlated (\mfig{fig:longmem}f). Correlations averaged over lags 5--15 also grow with the starting round (\mfig{fig:longmem}g).

Together, these decoder-independent observations show persistent detector correlations building up during the memory. Their temporal persistence and shared-qubit structure, together with the responses to reset and echo observed in the short-memory controls, are compatible with a persistent error component, potentially associated with leakage \cite{mcewen2021removing,miao2023overcoming} or environmental drift.

\newsection{Discussion}\label{sec:discussion}

We implemented rotated surface-code memories on the native square connectivity of an IBM Nighthawk r2 processor and examined how circuit design and hardware variation affect logical protection. Combining short readout, fused measurement and reset, and an echo on waiting data qubits reduced the logical error per round \(\varepsilon\) by a median of 29\,\% [24\,\%, 42\,\%] in Z memory and 73\,\% [69\,\%, 79\,\%] in X memory, relative to standard readout without reset or echo on the same qubits. With the other distance fixed at 3, increasing the protected distance from 3 to 6 approximately halved \(\varepsilon\) on the selected placements over 3--13 rounds. These results establish an improvement from circuit optimization and a distance advantage within the tested memory configurations.

The readout window provides a substantial opportunity for improving the syndrome-extraction circuit. The measurement instruction optimized for dynamic circuits shortens ancilla readout and lowers \(\varepsilon\) in our comparisons \cite{ibm2025dynamiccircuits}. Each qubit also has a dedicated dissipative reset element, making reset available during circuit execution \cite{ibm2026nighthawkr2}. Fused measurement and reset improves logical performance even when the total round duration is held fixed. The inserted-idle experiments further show that the cost of waiting depends on where it occurs in the circuit. An echo reduces X-memory errors, but extending the sequence beyond readout adds waiting time and eventually increases errors in both memories. These comparisons motivate evaluating readout, reset and decoupling together within the executed schedule. They complement observations of readout-idle costs and decoupling benefits on heavy-hex processors
\cite{vezvaee2026surface,harper2026characterising}.

Hardware variation also affects the benefit of increasing code size. Widening a patch introduces additional operations and increases the physical qubits used, and the protected-distance advantage varies with the duals distance. Among the tested superstabilizer layouts, excluding ancillas with poorer CZ couplers gives larger improvements than excluding ancillas whose main weakness is readout. This comparison motivates considering entangling gate quality alongside readout performance when choosing placements.

Decoder and noise-model choices provide another route to improved performance from the same measurement
records. However, fitting detector statistics does not necessarily improve every decoder (\snote{sn:decoders}), emphasizing the
importance of evaluating the model and decoder together. The complementary gap also distinguishes shots
with different observed failure probabilities, although its associated probability estimate is imperfectly
calibrated. Gap-based post-selection lowers conditional LEP by discarding shots. Reporting the retained
fraction alongside this improvement keeps its practical cost explicit and separates it from the
unconditional circuit and distance comparisons.

Longer memories expose a limitation of the short-memory improvement. The distance advantage weakens as
detection rates and persistent correlations build up during a shot. These changes are visible in the
syndrome records independent of decoding. Leakage-like errors are one possible contribution \cite{mcewen2021removing,miao2023overcoming} (\snote{sn:coincidence}), but the binary measurement records leave their source an open question. Measurements that distinguish higher qubit states could test this possibility, while further comparisons of reset and idle controls could establish which changes preserve logical protection over
longer times. Extending the observed improvement to longer memories and to patches with both distances
increased remains an experimental objective.

\newsection{Methods}\label{sec:methods}

\subsection{Device and circuit execution}\label{sec:m_device}
We implemented rotated surface-code memories on \phx{}, an IBM Nighthawk~r2 processor with 120 qubits on a
$12\times10$ square lattice, which fits the surface code naturally, so each patch was embedded without additional
SWAP gates.
A syndrome-extraction round comprised
four \rev{controlled-X (CX)} layers followed by ancilla readout and, where used, reset. During readout, the data qubits waited or
underwent dynamical decoupling. The CX gates were compiled into native CZ gates and single-qubit rotations.
\rev{Native CZ gates lasted \SI{68}{\nano\second}, and physical $X$ and $\sqrt{X}$ gates lasted \SI{40}{\nano\second}. $R_z$ rotations were virtual.}

On the selected placement, the four entangling layers and their associated single-qubit gates occupied
approximately \SI{0.45}{\micro\second}. Each qubit is paired with its own dissipative reset element, which
makes reset available in the middle of a circuit\cite{ibm2026nighthawkr2}. The mid-circuit measurement
instruction \ins{measure\_2}, optimized for dynamic circuits\cite{ibm2025dynamiccircuits}, reads a qubit in
\SI{1.14}{\micro\second} on this device, compared with \SI{1.94}{\micro\second} for the standard
\ins{measure} instruction. Each readout instruction also had a fused-reset variant, \ins{measure\_reset}
or \ins{measure\_reset\_2}, with the same duration. Rounds without reset or with fused reset lasted
approximately \SI{2.39}{\micro\second} with the long readout and \SI{1.59}{\micro\second} with the short
readout. A separate reset added \SI{0.40}{\micro\second}. The calibration of the qubits and couplers used is
listed in \snote{sn:device}.

\begin{figure*}[!tb]
\centering
\includegraphics[width=\linewidth]{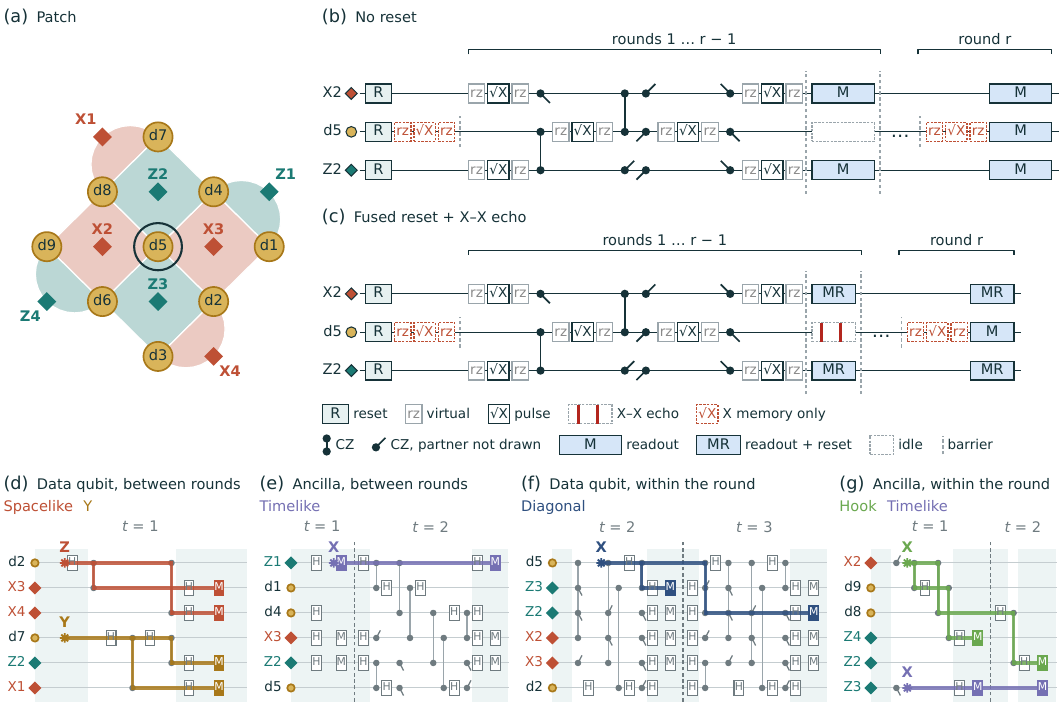}
\caption{\textbf{Executed circuits and detector patterns.}
The native schedules and representative fault patterns use the qubit labels in (a).
\pan{a} Distance-three patch, with d5 circled.
\pan{b},~\pan{c} Executed operations without reset and with fused short readout plus the nominal \SI{0.9}{\micro\second} characterization echo.
R, M and MR denote reset, measurement and fused measurement/reset.
\rev{Red dashed rotations apply only to X memory, and grey dashed lines are barriers.}
Connected dots denote CZ gates, and stubs indicate partners not shown. Schedules are not to scale.
\pan{d} Data faults between rounds produce space-like or mixed-family Y patterns.
\pan{e} Ancilla faults between rounds produce timelike patterns.
\pan{f} Data faults within a round produce diagonal patterns.
\pan{g} Ancilla faults within a round produce hook or timelike patterns.
Stars mark inserted faults.}
\label{fig:circuits}
\label{sfig:transpiled}
\end{figure*}

Circuits were constructed in Stim\cite{gidney2021stim}, converted to Qiskit\cite{javadiabhari2024quantum},
transpiled for the device and submitted through Qiskit Runtime. Barriers between circuit layers prevented
gates from being merged or cancelled across layer boundaries and preserved the intended basis during the
readout window. Executed schedules are shown in \mfig{fig:circuits}. Experimental conditions compared on a given placement
were submitted together in the same job (\snote{sn:drift}).
Unless stated otherwise, we prepared logical $|0\rangle$ and $|+\rangle$ states for the Z and X memories,
respectively, ran $r=3$, 5, 9 and 13 rounds, and acquired \num{20000} shots per circuit. Long-memory experiments
acquired \num{20000} shots at each of $r=1$, 3, 5, 9, 13, 21, 31, 45, 61, 81, 101, 125 and 150. All data qubits
were measured at the end of the circuit.

\subsection{Experimental protocols}\label{sec:m_protocols}
The lattice supported 96 distance-three placements without SWAP gates. On the scan day, one qubit had a
reported readout error of 50\,\%, no reported $T_1$ or $T_2$, and a reported CZ error probability of 1 on each
of its four couplers. We excluded placements containing that qubit and scanned the remaining 88. The selected
placement minimizes the combined Z- and X-memory scan LEP, \rev{decoded with MWPM on a calibration-derived detector error model. The second and third placements tested in the readout experiments are the next in this order that share no qubits with an earlier choice.} Rectangular patches, with $d_Z$ and $d_X$ the protected distances of the
Z and X memories, ran on up to three placements per distance pair $(d_Z,d_X)$, chosen by five-round scans or from
calibration data alone. The matched-duration reset comparison used the long readout.

Dynamical decoupling \cite{viola1998dynamical,viola1999dynamical,pokharel2018demonstration,tripathi2022delay,ezzell2023dynamical}
used an X--X sequence on each data qubit between full-patch barriers in the $r-1$ intermediate readout windows,
not during the final readout. \rev{Echo labels denote nominal delay budgets. The programmed delays were quantized to the hardware timing grid, and the two \SI{40}{\nano\second} X pulses added to the executed span.}
The baseline of the recipe comparison ran on one placement per distance pair.

For the spectator experiment, ancilla X3 of the selected placement served as the driver. It was prepared in
$|0\rangle$ or $|1\rangle$ and executed one readout instruction, or a delay of equal length as a control,
\rev{optionally followed by a separate reset. All qubits then waited 0, 0.4 or \SI{1.2}{\micro\second}.} Waits after
a separate reset were shortened by the reset duration, so that each reset circuit has a partner without reset
at equal elapsed time. We tested measurement alone and fused measurement and reset, each at both readout
durations. The four data qubits neighbouring the driver were prepared in $|+\rangle$ with a Hadamard gate and,
after the wait, measured in the X basis (a Hadamard before readout) or, in a separate circuit, in the Y basis
($S^\dagger$ and a Hadamard before readout). Their accumulated phase is
$\phi=\operatorname{atan2}(\langle Y\rangle,\langle X\rangle)$ (\snote{sn:spectator}).

\subsection{Detector correlations and error classification}\label{sec:m_correlations}
Measurement records were converted into detector and logical-observable parities defined by each executed
circuit. The circuit's detector error model identified the patterns produced by individual faults and
assigned detector pairs to space-like, timelike, diagonal, hook and mixed-family classes (\snote{sn:dem}).
This assignment accounted for the circuit schedule and reset scheme.

We estimated effective pairwise error probabilities from detector correlations\cite{google2021exponential}:
\begin{equation}
p_{ij} = \frac{1}{2} - \frac{1}{2}
\sqrt{1-\frac{4\left(\langle x_i x_j\rangle-\langle x_i\rangle\langle x_j\rangle\right)}
{1-2\langle x_i\rangle-2\langle x_j\rangle+4\langle x_i x_j\rangle}},
\label{eq:pij}
\end{equation}
where $x_i\in\{0,1\}$ indicates whether detector $i$ fired and averages run over shots. Under an
independent-error model, $p_{ij}$ estimates an effective probability of errors shared by the detector pair.
We averaged it over circuit-derived classes to compare error types\cite{google2023suppressing}.
Correlations between detectors associated with the same check at different rounds characterized persistent
signatures\cite{mcewen2021removing,miao2023overcoming,google2025dynamic}.

Without reset, bulk detectors compare an ancilla outcome with its outcome two rounds earlier\cite{geher2025reset}, and a single
fault can connect detectors up to two rounds apart. With reset, that reach is one round. We compared reset
schemes at matched offsets beyond their respective single-fault reach. In 13-round circuits, the long-tail
statistic averaged same-check $p_{ij}$ at offsets $k=2$ to 9 beyond that reach: $\Delta t=3$ to 10 with reset
and 4 to 11 without reset. This matching avoids counting the longer single-fault reach of the no-reset circuit
as evidence of an additional persistent component. Its intervals are described in \snote{sn:stats}.

\subsection{Noise models and logical-error estimation}\label{sec:m_noise}
For each executed circuit, we constructed a calibration-derived noise model from the run-day device snapshot and
archived operation list, reconstructing the schedule instruction by instruction for every qubit.
Measurement errors preceded readout with the calibrated error probability of the instruction executed on
that qubit. The same probability was used for errors after fused or separate reset. Each gate was followed
by depolarizing noise whose average error matched its calibration. Idle intervals carried Pauli-twirled
relaxation and dephasing noise derived from $T_1$ and $T_2$, with $T_2$ capped at $2T_1$.
\rev{Echo-protected intervals used the reported $T_2$, whereas unprotected intervals, including inserted delays, used an effective $T_2$ of 0.3 times that value, selected by decoding held-out shots from an earlier run.}

We decoded with MWPM\cite{higgott2025sparse} or, where stated, exact maximum-likelihood
decoding\cite{bravyi2014efficient,piveteau2024tensor}, \rev{both on the detector error model of this calibration-derived noise model. We call the former calibration-derived MWPM.} Comparisons with other decoders and fitted noise
models are described in \snote{sn:decoders}.

We defined $\mathrm{LEP}(r)$ as the fraction of shots whose decoded logical observable disagreed with
the prepared state after $r$ rounds. For short memories, we fitted\rev{\cite{google2021exponential}}
\begingroup\revcolor
\begin{equation}
1-2\,\mathrm{LEP}(r)=A(1-2\eps)^r
\label{eq:lep_fit}
\end{equation}
\endgroup
over $r=3$, 5, 9 and 13 by unweighted least squares of $\ln[1-2\,\mathrm{LEP}(r)]$ against $r$.
For intercept $a$ and slope $b$, $A=e^a$ and $\eps=(1-e^b)/2$. The fitted contrast amplitude allowed for
state-preparation and final-readout effects. Intervals came from binomial resampling of each round count
(\snote{sn:stats}).

At fixed other distance, we fitted $\ln\eps=\alpha+\beta d$ by unweighted least squares and defined the
two-distance-step suppression factor $\Lambda=e^{-2\beta}$. The statistical methods are described in
\snote{sn:stats}.

For long memories, we fitted all measured records with $r\ge3$ by binomial maximum likelihood to the
phenomenological cumulative-decay model
\begingroup\revcolor
\begin{align}
1-2\,\mathrm{LEP}(r)&=A\exp[-2I(r)],\label{eq:decay}\\
I(r)&=\eps_\infty r+(\eps_0-\eps_\infty)\tau\nonumber\\
&\quad{}\times(1-e^{-r/\tau}).\label{eq:decay_I}
\end{align}
\endgroup
Its parameter bounds, intervals and comparison with a constant-$\eps$ model are given in \snote{sn:stats}.

Windowed estimates were obtained separately from the endpoint contrasts $F(r)=1-2\,\mathrm{LEP}(r)$:
\begingroup\revcolor
\begin{equation}
\eps_{[r_1,r_2]}=
\frac{1}{2}\left[1-\left(\frac{F(r_2)}{F(r_1)}\right)^{1/(r_2-r_1)}\right].
\label{eq:window}
\end{equation}
\endgroup
We used windows 3--13, 13--31, 31--61, 61--101 and 101--150. These endpoint estimates eliminate a common
contrast amplitude and are distinct from fitting the global time-dependent model.

For confidence-based post-selection, we computed the complementary gap $g=|w_1-w_0|$, where $w_0$ and
$w_1$ are the minimum matching weights constrained to the two logical classes. This provides a matching-based
approximation to a log-likelihood ratio, with the associated failure-probability estimate $1/(1+e^g)$. Logical
failures were still evaluated using the ordinary MWPM prediction. Shots were ranked by largest gap or, for the
comparator, by lowest detection rate. At retained fractions $a=1$, 0.95, \ldots, 0.20, we kept the
$\mathrm{round}(an)$ highest-ranked of the $n$ shots.

\section*{Acknowledgements}\phantomsection\label{sec:ack}
This work was funded by the U.S. Department of Energy, Office of Science, National Quantum Information Science Research Center, Co-design Center for Quantum Advantage (C2QA) under contract number DE-SC0012704. This material is based upon work supported by the U.S. Department of Energy, Office of Science, National Quantum Information Science Research Centers, Quantum Science Center (QSC). The modelling and simulation work was supported by QSC, while experiments performed on IBM Quantum were supported by C2QA. This research was supported by PNNL’s Quantum Algorithms and Architecture for Domain Science (QuAADS) Laboratory Directed Research and Development (LDRD) Initiative. The Pacific Northwest National Laboratory is operated by Battelle for the U.S. Department of Energy under Contract No. DE-AC05-76RL01830. This research used resources of the Oak Ridge Leadership Computing Facility (OLCF), a DOE Office of Science User Facility supported under Contract No. DE-AC05-00OR22725. This research used resources of the National Energy Research Scientific Computing Center (NERSC), a U.S. Department of Energy Office of Science User Facility located at Lawrence Berkeley National Laboratory and operated under Contract No. DE-AC02-05CH11231, using NERSC awards DDR-ERCAP0038362, ASCRERCAP0037552, and DDR-ERCAP0038957. This research was supported in part by the National Science Foundation (NSF) under grant agreements 2329020 and 2343535.

\renewcommand{\bibsection}{\section*{References}}
\makeatletter\renewcommand{\@biblabel}[1]{#1.}\makeatother
\bibliographystyle{naturemag}
\widowpenalties=4 10000 10000 10000 0
\bibliography{refs}
\widowpenalties=0

\appendix
\renewcommand{\thesubsection}{\thesection\arabic{subsection}}
\makeatletter\renewcommand{\p@subsection}{}\makeatother   %

\section{Statistical methods}\label{sn:stats}

Quoted intervals are 95\,\% confidence intervals unless stated otherwise, and plotted LEP intervals are two-sided
Wilson intervals\cite{wilson1927probable}. As in ref.~\cite{google2025below}, $\eps$ is obtained from a linear fit of
$\ln[1-2\,\mathrm{LEP}(r)]$ against $r$, and $\Lambda$ from a linear fit of $\ln\eps$ against $d$. Their intervals are
95\,\% bootstrap percentiles, as in ref.~\cite{lacroix2025scaling}, here from a parametric bootstrap: the failures at each round count are redrawn from a binomial distribution with the
measured failure probability and the fits are repeated, 200 times for $\eps$ and 400 times for $\Lambda$ and for the
$\eps$ of \mfig{fig:device} and \mfig{fig:scaling}. Detection probabilities, $p_{ij}$ and
long-tail estimates use 200 bootstrap resamples of whole shots. \rev{Averages over placements or memories combine the individual half-widths in quadrature, divided by their number. Intervals on ratios divide opposite endpoints of the two marginal intervals. The interval on the median recipe gain resamples the eleven pair gains 400 times.} Long-memory pair
correlations are summarized by the mean and 1.96 standard errors across 22 patch--memory combinations, and the
cumulative-decay model is fitted by binomial maximum likelihood with bounded parameters and compared with a
constant-$\eps$ model through its deviance.

\section{Device, placements and circuits}\label{sn:app_device}

\subsection{Device parameters}\label{sn:device}

Table~\ref{tab:device} lists the reported calibration of the qubits and couplers used on each run day, from the
snapshot used to build that day's decoder model. Gate and readout durations are given in \methref{sec:m_device}.

\begin{table}[!tb]
\caption{\textbf{Device calibration on the run days.}
Median [range] over the qubits and CZ couplers of the executed memory circuits: the three placements (day 1), the
distance-scaling experiment (day 2) and the long memories (day 3). The day-3 $T_1$ and $T_2$ values date from
day 2. Errors are those reported for $\sqrt{X}$, CZ and the two readout instructions.}
\label{tab:device}
\begin{ruledtabular}
\footnotesize
\begin{tabular}{@{}lccc@{}}
 & Day 1 & Day 2 & Day 3 \\
\hline
$T_1$ (\si{\micro\second}) & 190 [66, 310] & 184 [26, 268] & 181 [26, 268] \\
$T_2$ (\si{\micro\second}) & 160 [12, 274] & 162 [4.6, 286] & 154 [4.6, 286] \\
$\sqrt{X}$ ($10^{-4}$) & 1.9 [1.2, 25] & 2.1 [0.98, 49] & 2.1 [0.98, 49] \\
CZ ($10^{-3}$) & 1.4 [0.87, 16] & 1.6 [0.80, 94] & 1.6 [0.78, 96] \\
\ins{measure} (\%) & 0.51 [0.23, 2.6] & 0.56 [0.20, 5.9] & 0.51 [0.15, 6.4] \\
\ins{measure\_2} (\%) & 0.44 [0.07, 8.7] & 0.46 [0.10, 8.6] & 0.42 [0.10, 8.8] \\
\end{tabular}
\end{ruledtabular}
\end{table}

\subsection{Placements: scan, selection and transposes}\label{sn:scan}\label{sn:transpose}

\begin{figure}[!tb]
\centering
\includegraphics[width=\linewidth]{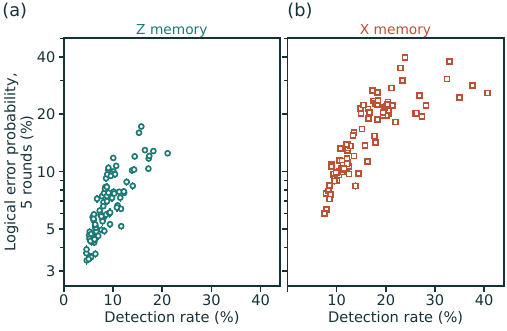}
\caption{\textbf{Detection and logical error across placements.}
Higher detection probabilities accompany higher LEP across the placement scan.
\pan{a},~\pan{b} LEP after five rounds versus the bulk detection probability of each memory's check family, for Z and X memory.
Each point represents one of 88 distance-three placements.
The scan uses standard readout without reset or echo.
Both panels share a logarithmic vertical scale.
Error bars are two-sided 95\,\% Wilson intervals.}
\label{fig:scan_det_lep}
\end{figure}

\rev{The scan conditions and the choice of the three distance-three placements are given in \methref{sec:m_protocols}. Each placement and memory has 20,000 or 40,000 shots.} The maps in \mfig{fig:device}b show the 44 placements with the illustrated
orientation. Across the scan, own-family detection probability correlates with the LEP from calibration-derived MWPM,
with Spearman coefficients of 0.856 [0.774, 0.907] in Z memory and 0.895 [0.840, 0.930] \rev{in X memory (\sfig{fig:scan_det_lep}). The intervals use 2,000 resamples of the 88 placements.} The correlation provides a
placement diagnostic without requiring a unique attribution of the observed detector activity to a particular error
process.

\begin{figure}[!tb]
\centering
\includegraphics[width=\linewidth]{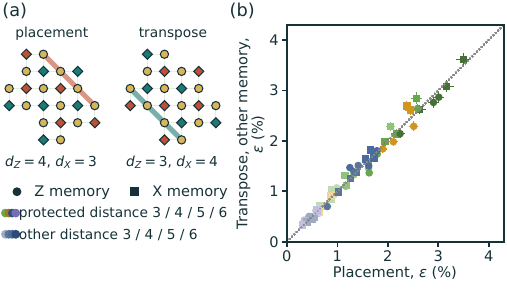}
\caption{\textbf{Transposed placements.}
Memories protecting the same physical direction have similar logical errors.
\pan{a} A placement and its transpose \rev{use the same data qubits while exchanging check roles. Highlighted paths have the same protected distance.}
\pan{b} $\eps$ for each placement versus that of the complementary memory on its transpose, for 60 recipe comparisons.
Colour identifies protected distance, shade the other distance and marker the original memory.
The dotted line marks equality.
Error bars are 95\,\% intervals from 200 binomial resamples of each round count.}
\label{sfig:transpose}
\end{figure}

The distance-scaling curves use, for each memory and distance pair, the full-patch placement with the lowest
$\eps$ under calibration-derived MWPM among the tested placements and their transposes. A transposed placement uses
the same physical qubits with the X- and Z-check roles exchanged (\sfig{sfig:transpose}a). The Z memory of one is
compared with the X memory of the other, so that the protected distance and physical direction are unchanged. The
comparison supports similar typical performance under the exchange, rather than exact equivalence of the noise in the
two memories (\sfig{sfig:transpose}b).

\subsection{Detector-pair classes}\label{sn:dem}\label{sn:round}

\begin{figure}[!tb]
\centering
\includegraphics[width=0.7\linewidth]{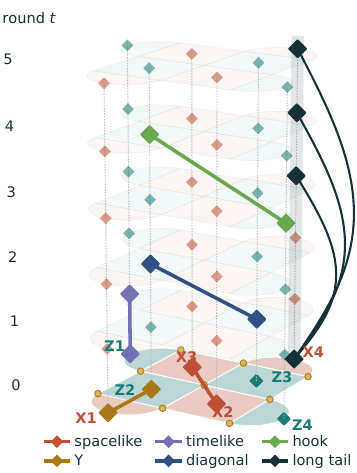}
\caption{\textbf{Detector pairs in space and time.}
The diagram distinguishes single-fault classes from correlations extending across several rounds.
The patch lies beneath successive detector layers, with vertical lines following each check through time.
Coloured links illustrate space-like, timelike, diagonal, hook and mixed-family Y pairs.
Dark arcs schematically represent persistent same-check correlations beyond the temporal reach of a single fault.}
\label{sfig:tail3d}
\end{figure}

The executed circuit (\mfig{fig:circuits}) determines which detector patterns a single fault can produce. We
construct these patterns with Stim \cite{gidney2021stim} from each executed schedule and use them to classify measured
detector pairs (\mfig{fig:circuits}d--g; \sfig{sfig:tail3d}). A data-qubit fault between rounds can produce a
space-like pair in the same round. A fault between interactions with two checks can instead produce a diagonal pair
separated in both space and time. Mixed-family pairs can arise from Y-type faults. An ancilla fault can produce a
timelike pair or propagate through later entangling gates to produce a hook-class pair \cite{tomita2014low}. \rev{These are circuit-derived signatures, and a measured correlation does not by itself identify a unique microscopic fault.} The
single-fault reach, which sets the long-tail offsets, depends on the reset scheme (\methref{sec:m_correlations}).

\subsection{Superstabilizer layouts}\label{sn:superstab}

Defect-adapted surface codes use modified check measurements and superstabilizers to avoid unavailable
components \cite{auger2017fault,nagayama2017surface}. All four excluded qubits---q24, q55, q62 and q88---serve as
ancillas in the full patches used for these comparisons, including transposed placements, and their X- or Z-check
roles can vary with placement. Each adapted circuit excludes the selected ancilla and its entangling gates.

In the \(4\times5\) example in \mfig{fig:superstab}a, two neighbouring ancillas measure their ordinary X checks
on alternating rounds and weight-two Z-gauge operators on the intervening recovery rounds. The product of the
Z-gauge outcomes reconstructs the omitted Z-check information. The affected X-check gauges are combined into an
X superstabilizer, while unaffected checks continue to be measured every round. The corresponding detector
definitions account for this two-round cadence. No syndrome-extraction gate uses the excluded ancilla or its
couplers. This alternating use of neighbouring ancillas follows the ancilla-repurposing construction of
Leroux et al.\ \cite{leroux2025snakes}. Both layouts use the recipe on the same placement and memory basis (\methref{sec:m_protocols}).

\subsection{Device stability within a job}\label{sn:drift}

IBM executes the circuits of a job together, iterating over all of them for each shot\cite{ibm_repetition_rate}.
The same shot index of every circuit in a job therefore falls in the same stretch of the job, which lasts 40~s to
4~minutes, and a change of the device during the job is shared by all its circuits. For the same reason, a circuit
placed last in a job, such as the repeated $r=21$ circuit of the long-memory jobs, is a replicate rather than an
end-of-job measurement. The returned shot order is consistent with chronological order: neighbouring blocks of
\num{2000} shots of a circuit have more similar detection rates than distant blocks, which none of 300 random
reorderings reproduces, and two circuits of the same job share the same block-by-block detection-rate profile
(correlation $+0.29$ in the long-memory jobs), whereas circuits of different jobs do not ($-0.01$).

\begin{figure}[!tb]
\centering
\includegraphics[width=\linewidth]{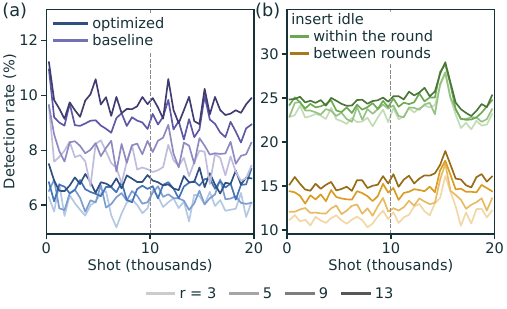}
\caption{\textbf{Device drift within a job.}
A change of the device during a job is shared by all its circuits.
Bulk detection rate of single Z-memory circuits against shot number, \rev{in bins of 500 shots, one line per circuit, shaded by the number of rounds $r$. The dashed line marks half of the shots.}
\pan{a} A job of the distance-scaling experiment (the 3\,$\times$\,3 patch of \mfig{fig:device}c), optimized and
baseline circuits: flat apart from the first 500 shots of the baseline circuits.
\pan{b} A job of the inserted-idle experiment (\mfig{fig:idle}): every circuit peaks between shots \num{14500} and
\num{15500}, 3.2--6.0 points above its median.}
\label{sfig:drift}
\end{figure}

\sfig{sfig:drift} shows such shared drift. In the job of \sfig{sfig:drift}b, all 80 circuits peak between shots
\num{14500} and \num{15500}, \rev{3.2--6.0 points above their median, and then return. In the job of \sfig{sfig:drift}a, by contrast, every circuit stays at its level apart from the first 500 shots of the baseline circuits.} Across all \num{2364}
decoded memory circuits of the three run days, the logical error of the first and second halves of each circuit's
shots differs more than random splits of the same shots allow, strongly on the three placements and in the
echo-length experiment ($p<0.001$) and weakly in the others ($p=0.002$--0.05). The error per round shows no systematic
change through a job (median ratio of the two halves 0.98--1.01).

\section{Readout duration and reset}\label{sn:app_window}

\subsection{Readout-window error budget}\label{sn:reset_budget}

\begin{figure}[!tb]
\centering
\includegraphics[width=\linewidth]{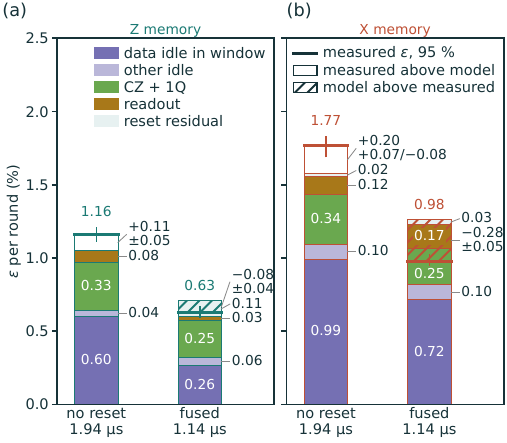}
\caption{\textbf{Readout-window error budget.}
The budget underpredicts long-readout $\eps$ and overpredicts fused-short-readout $\eps$.
\pan{a},~\pan{b} Model contributions and measured $\eps$ in Z and X memory on the selected placement, without echo.
White and hatched regions show measured and model excess, respectively.
Measured-minus-model differences are $+0.11$ [0.06, 0.16] and $-0.08$ [$-0.12$, $-0.04$] percentage points in Z memory, and $+0.20$ [0.11, 0.26] and $-0.28$ [$-0.33$, $-0.23$] in X memory, for long and fused-short readout.
These 95\,\% intervals include hardware shot uncertainty only.}
\label{sfig:reset_budget}
\end{figure}

The budget in \sfig{sfig:reset_budget} compares the long-readout circuit without reset with the short
fused-readout circuit on the selected placement, neither with an echo. As in ref.~\cite{google2025below}, each noise
group's contribution is its sensitivity in simulation: each of six groups is scaled to 0.75 and 1.25 of its baseline,
\num{50000} shots are simulated per setting and decoded with MWPM, and the derivative of the fitted $\eps$ at the
baseline, normalized over the groups, apportions the baseline $\eps$. The simulated model is the calibration-derived
model with the data-qubit readout-window channels and the ancilla readout and fused-reset errors replaced by same-day
measurements: window dephasing from Ramsey-contrast decay and population loss from the $|1\rangle$-loss slope on six
data qubits, with their median channel on the other three.

Readout idling is the largest modelled contribution: in the long-readout circuit it accounts for 57\,\% of the
modelled $\eps_Z$ and 63\,\% of the modelled $\eps_X$. The model predicts reductions of 0.34 percentage points in Z
memory and 0.32 points in X memory between the two circuits, against measured point-estimate reductions of 0.53 and
0.79 points. The residuals have both signs, so they cannot be interpreted as a direct measurement of a single omitted
error process.

\subsection{Idle-decoherence prediction for space-like pairs}\label{sn:idle_pred}

The idle-only comparison in \mfig{fig:window}d treats the shorter wait as a Pauli-twirled relaxation and dephasing
channel on the data qubit shared by each space-like detector pair. Over an idle of duration \(t\), this channel flips
the qubit with probability \(P_{\mathrm{bit}}(t)=(1-e^{-t/T_1})/2\) and dephases it with probability
\(P_{\mathrm{phase}}(t)=(1-e^{-t/T_2})/2\). For each pair, we evaluate the reduction between 1.94 and
\SI{1.14}{\micro\second} with the reported \(T_1\) of the shared data qubit for bit-flip-associated (Z-check) pairs and
its reported \(T_2\) for phase-associated (X-check) pairs. Unlike the budget, this reference takes the reported
decoherence times directly. Across 54 comparisons per error class, the measured median reduction matches the
prediction for bit-flip-associated pairs but exceeds it for phase-associated pairs, which supports an additional
phase-sensitive contribution.

\subsection{Detection probability by reset scheme and readout duration}\label{sn:readout_det}

\begin{figure*}[!tb]
\centering
\includegraphics[width=\linewidth]{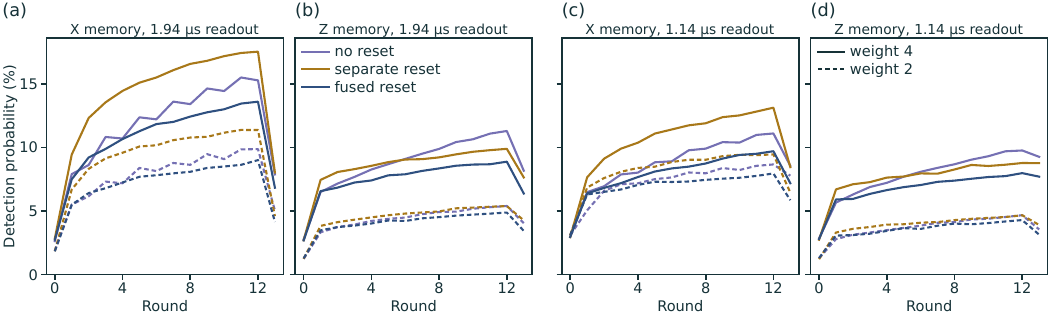}
\caption{\textbf{Detection probability by reset scheme and readout duration.}
Without reset, detection probability rises fastest through the memory.
Mean detection probability of the memory's own checks against the round, averaged over the three placements, for the
natural-duration circuits of \mfig{fig:window}.
\pan{a} X checks in X memory and \pan{b} Z checks in Z memory with long (\SI{1.94}{\micro\second}) readout;
\pan{c},~\pan{d} the same with short (\SI{1.14}{\micro\second}) readout.
Colours: no reset, separate reset and fused reset; solid lines: weight-4 checks; dashed lines: weight-2 checks.
13-round circuits, \num{20000} shots each. The last point compares the final data-qubit readout.}
\label{sfig:readout_det}
\end{figure*}

Without reset, the detection probability of each memory's own weight-4 checks rises fastest through the memory
(\sfig{sfig:readout_det}). A linear fit over rounds 2--12 gives a rise of 0.42 percentage points per round in Z
memory with long readout, \rev{against 0.18 with separate and 0.20 with fused reset. In X memory the margin is smaller: 0.63 against 0.49 and 0.43.} Short readout shows the same ordering (Z: 0.35 against 0.17 and 0.19; X: 0.40 against 0.38 and 0.29).
Averaged over these rounds, shorter readout lowers the detection probability for every reset scheme, and fused reset
gives the lowest value in every memory and readout duration. Separate reset gives the highest X-check detection
probability in X memory, 15.7\,\% against 12.9\,\% without reset at long readout and 11.5\,\% against 9.4\,\% at short
readout, in the same direction as its larger read-1 neighbour excess (\snote{sn:conditioned}).

\subsection{Spectator phase and excited-state exposure}\label{sn:spectator}\label{sn:exposure}

\begin{figure*}[!tb]
\centering
\includegraphics[width=\linewidth]{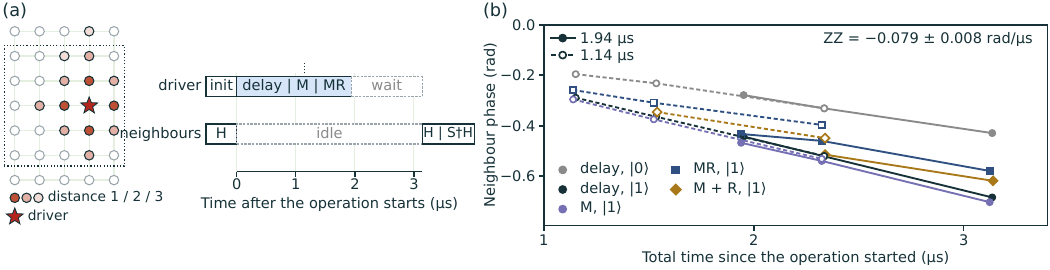}
\caption{\textbf{Neighbour phase during ancilla readout.}
Neighbour phase depends on the driver's state and readout/reset instruction.
\pan{a} Driver location, neighbouring qubits and spectator sequence.
\pan{b} Mean phase of the four nearest neighbours versus elapsed time, with the driver delayed in $|0\rangle$ or $|1\rangle$, or measured in $|1\rangle$ with no, fused or separate reset.
Solid filled and dashed open series denote long and short readout.
The annotation gives the fitted state-dependent phase slope, labelled ZZ.
Error bars are 95\,\% intervals from 200 Poisson-weighted whole-shot resamples (\snote{sn:spectator}).}
\label{sfig:spectator}
\end{figure*}

The spectator protocol is described in \methref{sec:m_protocols} (\sfig{sfig:spectator}a). The difference between the two
driver-state phase slopes gives a mean effective coupling

\begingroup\revcolor
\begin{equation}
\zeta=-0.079\;[-0.087,-0.071]\ {\rm rad}\,\mu{\rm s}^{-1},
\label{eq:zeta}
\end{equation}
\endgroup

\rev{where the sign follows the phase convention of the measurement (\sfig{sfig:spectator}b). The model below uses its magnitude.} \rev{The mean phase averages the four neighbouring data qubits. Its intervals use 200 Poisson-weighted whole-shot resamples, preserving correlations between simultaneously measured neighbours.} The fitted-coupling
interval combines the four qubits' half-widths in quadrature, divided by four.

From the instruction-dependent phase response we infer an effective time the driver spends excited before readout
ends.
At the elapsed times of the matched-duration comparison, these exposure intervals are 2.40 [2.18, 2.72]
\si{\micro\second} without reset, 2.11 [1.91, 2.41] \si{\micro\second} with separate reset and 1.48 [1.26, 1.73]
\si{\micro\second} with fused reset. They are inferred from phase accumulation, not measured as an excited-state
residence time, \rev{and come from one driver ancilla. Applied to other ancillas, they therefore give a reference prediction rather than a complete calibration of the patch.}

An uninterrupted excited interval \(t_{\mathrm{exc}}\) produces a phase angle \(\zeta t_{\mathrm{exc}}\), with
effective single-qubit phase-error probability

\begingroup\revcolor
\begin{equation}
p_\phi(t_{\mathrm{exc}})
=\sin^2\!\left(\frac{\zeta t_{\mathrm{exc}}}{2}\right).
\label{eq:pphi}
\end{equation}
\endgroup

Using the recorded read-1 fraction \(f\) as the exposure weight gives, in the weak-phase limit,

\begingroup\revcolor
\begin{equation}
\overline p_\phi
\approx f\,\frac{\zeta^2t_{\mathrm{exc}}^2}{4},
\label{eq:pphi_mean}
\end{equation}
\endgroup

so the relevant weighting is \(f t_{\mathrm{exc}}^2\), rather than the total-time proxy \(f t_{\mathrm{exc}}\). For a
stable check value of 1, a no-reset ancilla record alternates between 0 and 1, whereas with reset the same check value
can produce a read-1 outcome in every round. The observed bulk read-1 fractions are approximately 35\,\% without reset
and \rev{49\,\% with either reset. The model uses these recorded fractions as weights, not as population measurements.}

With these fractions and exposure intervals, the total-time proxy \(f t_{\mathrm{exc}}\) is 0.84, 1.04 and
\SI{0.73}{\micro\second} without, with separate and with fused reset, but the relative squared-exposure weighting
\(f t_{\mathrm{exc}}^2\) is 1.00, 1.08 and 0.53. \rev{Separate reset shortens each excited interval, but its larger read-1 fraction offsets that reduction. Fused reset shortens the interval more strongly.} The small-phase two-neighbour
approximation \(2f\sin^2(\zeta t_{\mathrm{exc}}/2)\) predicts weighted neighbour excesses of 0.63, 0.68 and 0.34
percentage points, computed from the unrounded fractions and coupling. Uncertainty in the coupling and exposure
estimates is not propagated into these predictions.

\subsection{Neighbouring X checks after an ancilla reads 1}\label{sn:conditioned}

The read-1 neighbour comparison in \mfig{fig:zz}c uses the bulk Z-check ancillas Z2 and Z3 in 13-round X-memory
records from the three placements and the three equal-duration reset schemes. \rev{The read-1 condition retains histories 101 without reset and 111 with reset.} The read-0 reference retains 000 for
every reset scheme. Histories are taken at zero-based round indices $t=2,\ldots,11$. We compare next-round detection
probabilities for X-checks sharing two data qubits with the driver and subtract the corresponding contrast for
checks sharing no data qubits. Counts are pooled over rounds, drivers and placements. The test is restricted to bulk
ancillas because boundary checks have an additional stabilizer constraint: in an idealized equal-coupling model,
equal phase rotations on the two data qubits of an odd-parity boundary check can cancel in the relevant detection
probability.

Before weighting, the read-1 neighbour excess is 3.49 [3.36, 3.62] percentage points without reset, 2.93
[2.82, 3.03] with separate reset and 1.54 [1.44, 1.64] with fused reset (\mfig{fig:zz}c). Weighted by the driver's
read-1 fraction over all eligible rounds, it is 1.23 [1.18, 1.27], 1.44 [1.39, 1.49] and 0.75 [0.71, 0.80]
percentage points, respectively. These 95\,\% intervals treat that fraction as fixed and use a normal approximation
to binomial counting uncertainty, treating the contributing detector samples as independent. The weighted excess is
a diagnostic, not a direct measurement of a microscopic phase-error probability.

The measured weighted excesses follow the ordering predicted in \snote{sn:exposure} (0.63, 0.68 and 0.34 points) and
are approximately twice the predictions for every reset scheme. The exposure model therefore reproduces the ordering of the measured
excesses. The retained histories differ in check value as well as
recorded ancilla state, so their contrast \rev{need not arise entirely from excitation-dependent phase. Variation in the coupling between ancillas and other phase contributions are further possible sources of the discrepancy.} \rev{This selection of rounds is used only for this comparison, whereas the logical-error comparisons retain all shots.}

\section{Persistent detector correlations}\label{sn:leak}

We examine the persistent detector correlations first in time and then in space. \sfig{sfig:pairmaps} and
\sfig{sfig:leaksig} \rev{use the same 13-round records of one placement. \sfig{sfig:pairmaps} follows individual check pairs round by round, showing how long a pair of detectors stays correlated and how ancilla reset changes it. \sfig{sfig:leaksig} summarizes, for every pair of checks,} how often both are persistently active in the same shot, pooled
over both memories and readout durations.

\subsection{Time profile of persistent correlations}\label{sn:reset_tail}\label{sn:pairmaps}

\begin{figure*}[!tb]
\centering
\includegraphics[width=\linewidth]{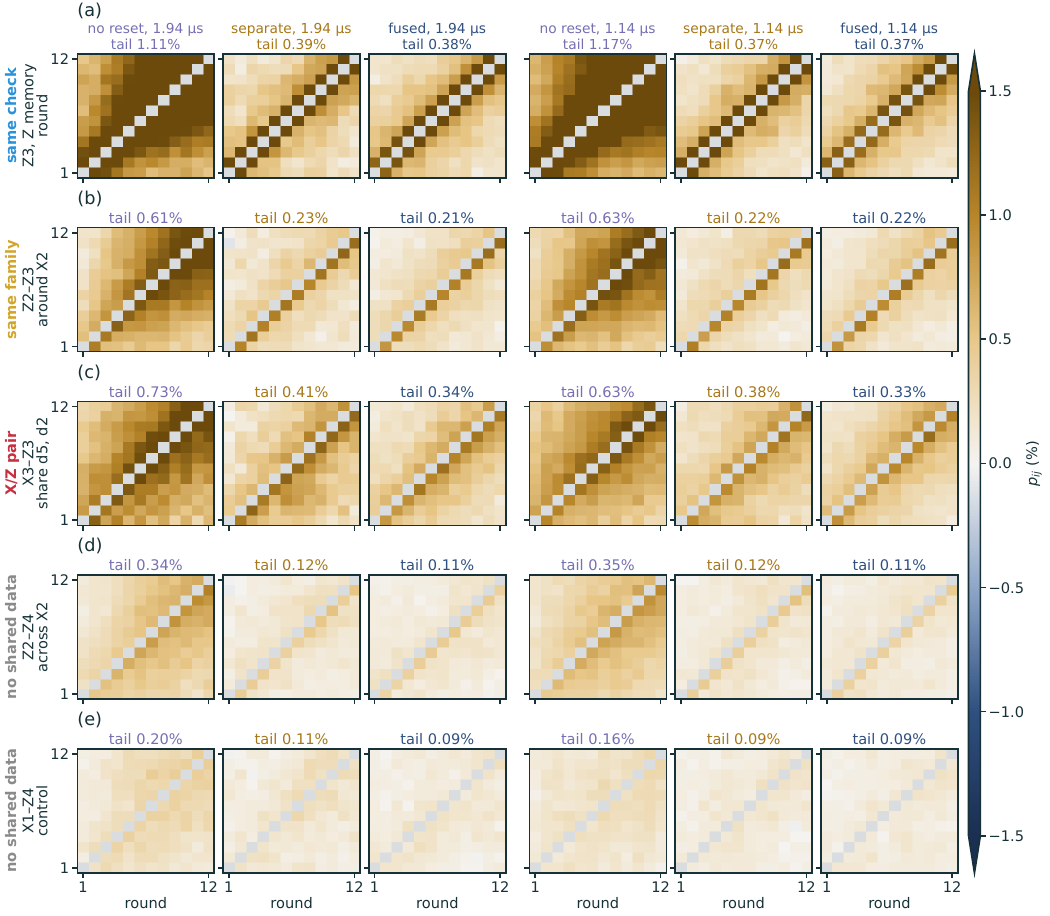}
\caption{\textbf{Correlation maps by pair type.}
Reset-dependent changes differ among detector-pair classes.
\pan{a} Same check: Z3, Z memory.
\pan{b} Z2--Z3 around X2.
\pan{c} X3--Z3, sharing d5 and d2.
\pan{d} Z2--Z4 across X2.
\pan{e} X1--Z4, the distant control.
Columns show no, separate and fused reset at long readout, then short readout.
\rev{Axes give detector rounds and titles the long-tail values.}
\rev{All panels use the same placement as \sfig{sfig:leaksig}, and (b)--(e) average both memories.}}
\label{sfig:pairmaps}
\end{figure*}

The long-tail comparison in \mfig{fig:zz}d--f uses the matched-duration circuits and the temporal offsets defined in \methref{sec:m_correlations}. Reset reduces some check-pair correlations strongly while leaving others substantial. For the Z3 check shown in \sfig{sfig:pairmaps}a (Z memory), the same-check tail changes from 1.04\,\% [0.96\,\%,
1.11\,\%] without reset to 0.39\,\% [0.35\,\%, 0.44\,\%] with separate reset and 0.41\,\% [0.37\,\%,
0.44\,\%] with fused reset.
In contrast,
X3 of the second placement in X memory changes from 3.24\,\% [3.08\,\%, 3.47\,\%] to 3.47\,\% [3.27\,\%, 3.67\,\%]
and 3.13\,\% [2.94\,\%, 3.30\,\%], respectively.

The correlation maps in \sfig{sfig:pairmaps} show these responses round by round. Each map gives the effective pair
probability \(p_{ij}\) as a function of the two detector rounds, for no, separate and fused reset at long readout and
then at short readout. Panel (a) shows the same-check map of Z3 in Z memory. In these natural-duration circuits its long tail is 1.11\,\% without reset and 0.39\,\% and 0.38\,\% with separate and
fused reset at long readout, and 1.17\,\%, 0.37\,\% and 0.37\,\% at short readout.
Panels (b)--(e) average both memories for individual check pairs: Z2--Z3 around X2; X3--Z3, sharing data qubits d5
and d2; Z2--Z4, which share no data qubit but lie across X2; and X1--Z4, a further pair without shared data qubits.
The same-family Z2--Z3 and Z2--Z4 tails decrease with either reset, the mixed-family X3--Z3 pair retains a larger
tail, and the X1--Z4 control is less strongly correlated. The maps use the natural-duration circuits, whereas
\mfig{fig:zz} uses padded matched-duration circuits, so absolute tail values should be compared within each design.
X-check tails also decrease with the data-qubit echo (\mfig{fig:echo}d; \sfig{sfig:echo_p35}d).

Separate reset can substantially reduce persistent correlations without improving X-memory $\eps$, so long-tail
suppression alone does not explain the logical-performance comparison.

\subsection{Spatial pattern of persistent correlations}\label{sn:coincidence}

We examine simultaneous activity of check pairs in 13-round records to characterize persistent correlations across a
patch. Each shot is divided into two windows of six rounds. A check is called elevated when it fires in at least
three rounds of a window. For each pair of checks, we compare the frequency of simultaneous elevation with a reference
obtained by independently permuting the shot order of each check's complete time record, averaged over three such
permutations. This preserves each check's firing rate and temporal structure while removing correlations between
checks.

\begin{figure*}[!tb]
\centering
\includegraphics[width=\linewidth]{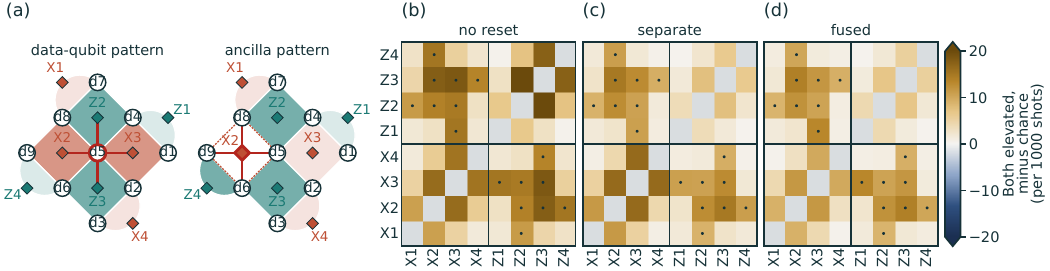}
\caption{\textbf{Simultaneous check activity.}
The pattern of excess simultaneous activity changes with reset scheme.
\pan{a} Candidate check patterns for a persistent error associated with a data qubit or an ancilla.
\pan{b}--\pan{d} Frequency with which two checks each fire in at least three of six rounds, after subtracting a shuffled-shot reference, without reset, with separate reset and with fused reset.
Results pool both memories and readout durations on the same placement as \sfig{sfig:pairmaps}.
\rev{Dots identify mixed-family pairs sharing a data qubit, and grey marks the diagonal.}}
\label{sfig:leaksig}
\end{figure*}

The schematics in \sfig{sfig:leaksig}a illustrate candidate patterns associated with a persistent error near a
data qubit or an ancilla. The maps in \sfig{sfig:leaksig}b--d show the excess simultaneous activity without reset, with separate reset and with fused reset, pooling both memories and readout durations.
Dots identify X/Z-check pairs sharing a data qubit. Checks sharing data qubits show persistent simultaneous activity
under all three reset schemes, while some same-family pairs around the X2 ancilla are more strongly reduced by reset.
\rev{These observations are consistent with persistent contributions that respond differently to ancilla reset. The schematic patterns provide possible interpretations of those responses.}

\subsection{Detection probability of single checks in long memories}\label{sn:longcheck}

\begin{figure*}[!tb]
\centering
\includegraphics[width=\linewidth]{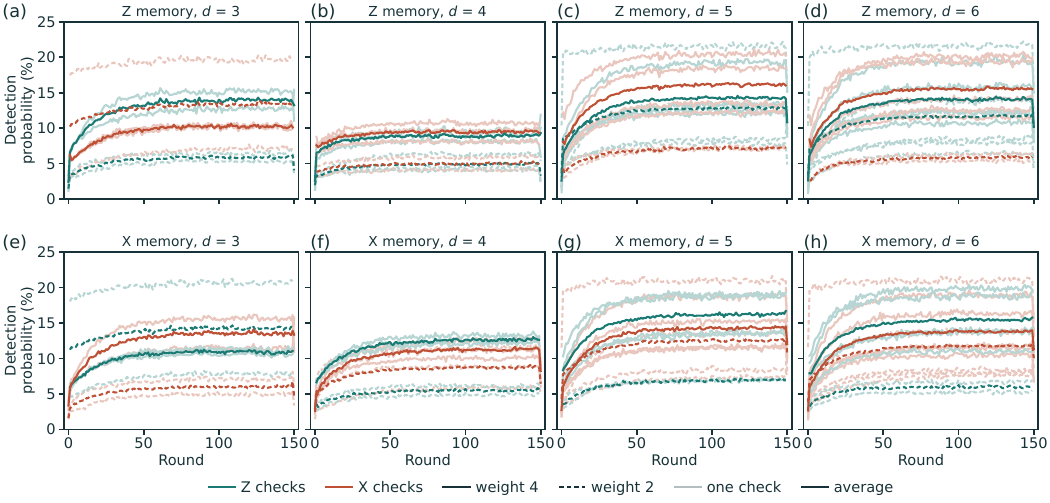}
\caption{\textbf{Detection probability of single checks in long memories.}
The rise of the detection probability in \mfig{fig:longmem}e is shared by nearly all checks.
\pan{a}--\pan{d} Z memory and \pan{e}--\pan{h} X memory, for protected distances 3--6 at other distance 3: detection
probability of every check against the round in the 150-round circuit of each patch.
Light lines: single checks; dark lines: averages of the weight-4 (solid) and weight-2 (dashed) checks of each type.
Z checks teal, X checks rust.
Optimized circuit, \num{20000} shots each.}
\label{sfig:longcheck}
\end{figure*}

\sfig{sfig:longcheck} shows the detection probability of \mfig{fig:longmem}e for each single check, for the 150-round
memory of each patch. Between rounds 3--13 and 101--149, the average of the weight-4 checks of each type rises by a
factor of 1.47--1.65 in every patch except the distance-four Z-memory patch (1.18--1.21). The rise is not carried by a
few checks: across the 100 checks the median factor is 1.54, and 94\,\% of the checks rise by more than 10\,\%. In six
of the eight patches a single check starts near 20\,\% (18--21\,\%) and rises little, to 20--22\,\%. A detection
probability that grows in nearly every check over tens of rounds is consistent with a persistent error component
that accumulates during the memory, rather than with a few defective qubits. A similar rise in the detection
probability of all detectors was observed in surface-code memories without data-qubit leakage removal, attributed
to leakage accumulation, \rev{and removed by data-qubit leakage removal (ref.~\citenum{google2025below}, Fig.~S11). Our circuits contain none.}

\section{Idle placement and dynamical decoupling}\label{sn:app_idle}

\subsection{Inserted idles within and between rounds}\label{sn:idle}

We insert pulse-free delays of 0.48, 0.96 and \SI{1.92}{\micro\second} on every qubit of the selected placement,
either after the second entangling layer or after ancilla readout. Each circuit is decoded using the calibration-derived
model of its executed schedule, including the inserted delays.

\begin{figure*}[!tb]
\centering
\includegraphics[width=\linewidth]{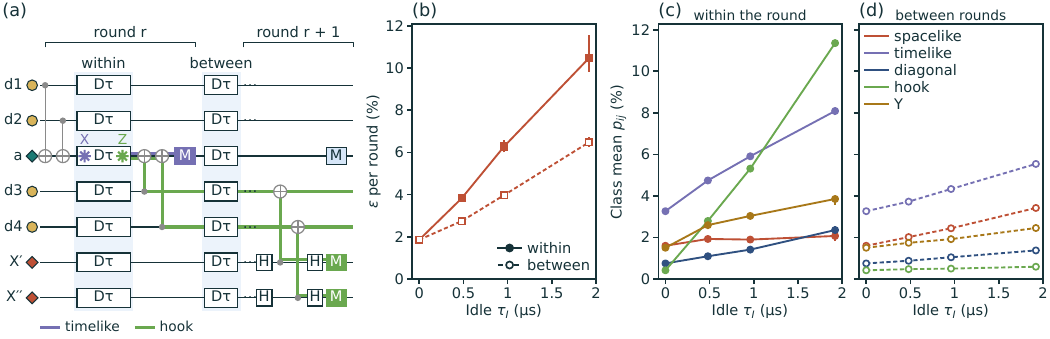}
\caption{\textbf{Idle placement in X memory.}
An inserted delay produces a larger logical-error increase within the round than between rounds.
\pan{a} Delay locations and illustrative ancilla-fault propagation.
\pan{b} $\eps$ versus inserted delay: solid filled series denote within-round insertion and dashed open series denote between-round insertion.
\pan{c},~\pan{d} Corresponding class-averaged $p_{ij}$, using a common scale.
Delays contain no pulses and act on every qubit of the selected placement.
Error bars are 95\,\% intervals from 200 binomial resamples of each round count (b) and from 200 whole-shot bootstrap resamples (c,d).}
\label{sfig:idle_x}
\end{figure*}

The X-memory measurements in \sfig{sfig:idle_x} reproduce the ordering observed in Z memory. For a
\SI{1.92}{\micro\second} delay, \(\varepsilon_X\) rises from 1.86\,\% [1.78\,\%, 1.94\,\%] without the added
delay to 10.47\,\% [9.82\,\%, 11.59\,\%] within the round, compared with 6.48\,\% [6.25\,\%, 6.73\,\%] between
rounds (\sfig{sfig:idle_x}b). The corresponding hook-class \(p_{ij}\) is 11.36\,\% [11.24\,\%, 11.48\,\%] within
the round and 0.59\,\% [0.52\,\%, 0.66\,\%] between rounds (\sfig{sfig:idle_x}c,d).

\subsection{Echo on a second placement}\label{sn:echo_p35}

\begin{figure*}[!tb]
\centering
\includegraphics[width=\linewidth]{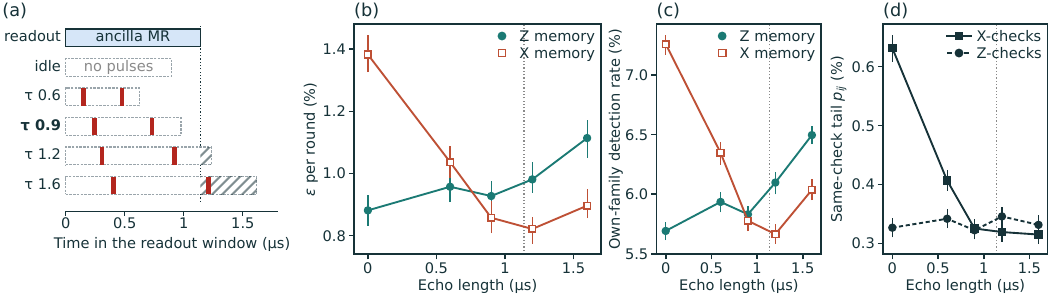}
\caption{\textbf{Readout echo on a second placement.}
As on the selected placement, the echo lowers X-memory error.
\rev{\pan{a} Echo schedules during fused short readout, with hatched segments beyond the readout.}
\pan{b} $\eps$ versus nominal echo length.
\pan{c} Detection probability of each memory's check family.
\pan{d} Same-check long-tail $p_{ij}$ for X- and Z-checks, averaged over both memories.
Zero denotes the no-echo control, and the dotted line marks the readout duration.
The longest sequences extend the round.
Error bars are 95\,\% intervals from 200 binomial resamples of each round count (b) \rev{or from 200 whole-shot bootstrap resamples (c,d). In (d), the two memories' half-widths are combined in quadrature (\snote{sn:stats}).}}
\label{sfig:echo_p35}
\end{figure*}

\sfig{sfig:echo_p35} repeats the echo-length sweep of \mfig{fig:echo} on the second placement. Relative to the no-echo
control, the nominal \SI{0.9}{\micro\second} echo reduces \(\varepsilon_X\) by 38.0\,\%, against 21.4\,\% on the
selected placement, while the \(\varepsilon_Z\) point estimate increases by 5.2\,\% (2.6\,\% on the selected
placement), with overlapping 95\,\% intervals on both placements (\sfig{sfig:echo_p35}b). The Z-check detection
probability increases by 2.4\,\% (4.7\,\% on the selected placement), and the X-check long tail decreases from
0.63\,\% [0.61\,\%, 0.65\,\%] to 0.33\,\% [0.31\,\%, 0.34\,\%], a reduction of approximately 48\,\%
(\sfig{sfig:echo_p35}c,d).

On the selected placement, increasing the nominal echo duration from 1.2 to \SI{1.6}{\micro\second} raises
\(\varepsilon_Z\) from 0.745\,\% [0.698\,\%, 0.791\,\%] to 0.893\,\% [0.844\,\%, 0.946\,\%] and \(\varepsilon_X\)
from 0.723\,\% [0.679\,\%, 0.770\,\%] \rev{to 0.811\,\% [0.771\,\%, 0.857\,\%]. The X-memory error still remains below its no-echo value.}

\subsection{Repeated readout windows with dynamical decoupling}\label{sn:justdd}

\begin{figure*}[!tb]
\centering
\includegraphics[width=\linewidth]{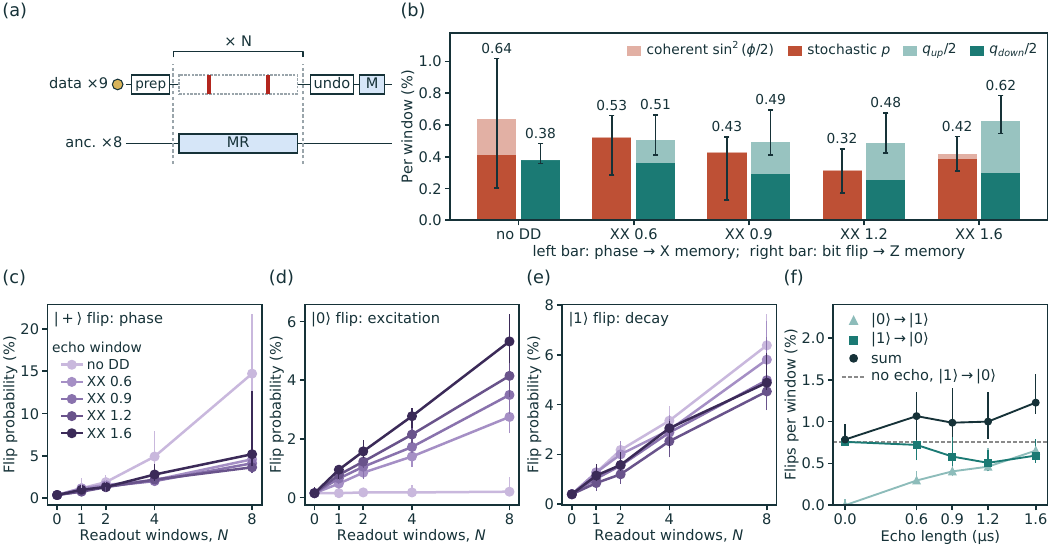}
\caption{\textbf{Echoes outside the code.}
Echoes reduce phase-sensitive flips while adding excitation.
\pan{a} Nine data qubits experience repeated readout windows without syndrome-extraction gates.
\pan{b} Model-based per-window phase and bit-flip contributions.
\pan{c}--\pan{e} Flip probabilities from $|+\rangle$, $|0\rangle$ and $|1\rangle$ versus window count.
\pan{f} Excitation, decay and their sum \rev{versus nominal echo length, with the dashed line marking decay without echo.}
\rev{Bars combine component medians and the other panels qubit medians.}
\rev{Error bars, shown in (b) for the total only, are 95\,\% intervals from \num{4000} resamples of the nine data qubits.}}
\label{sfig:justdd}
\end{figure*}

To examine the echo separately from syndrome extraction, we repeat only the readout window of a round, without
its entangling layers (\sfig{sfig:justdd}a). The nine data qubits of a distance-three patch are prepared in
\(|0\rangle\), in \(|1\rangle\), or in an equal superposition prepared with a \(\sqrt{X}\) pulse and labelled
\(|+\rangle\). They then pass through \(N = 1\), 2, 4 or 8 readout windows. In each window, the eight ancillas
execute \ins{measure\_reset\_2}, as in a fused-reset round, while the data qubits either wait for the length of the
window or \rev{run the same X--X sequence as in the code (\methref{sec:m_protocols}). Sequences longer than the \SI{1.14}{\micro\second} window extend it.} Barriers on all qubits separate consecutive windows. No CZ gates are applied, so the data and
ancilla qubits are never entangled, no detectors are formed and no decoding is involved. After the last window, the
preparation is undone and the data qubits are measured with \ins{measure\_2}. The flip probability is the fraction of
shots in which a data qubit \rev{does not return to \(|0\rangle\). Circuits with \(N = 0\) give the state-preparation-and-measurement floor.} Each circuit uses 4,000 shots. Because the ancillas remain close to
\(|0\rangle\), these windows do not reproduce the dependence of the data-qubit phase on the ancilla state
(\snote{sn:spectator}).

After eight windows, the median flip probability from \(|+\rangle\) decreases from 14.7\,\% [7.4\,\%, 21.8\,\%]
without an echo to 4.1\,\% [3.5\,\%, 5.0\,\%] with the nominal \SI{0.9}{\micro\second} echo. The corresponding
probability from \(|0\rangle\) increases from 0.2\,\% [0.1\,\%, 0.7\,\%] to 3.5\,\% [2.9\,\%, 4.8\,\%]
(\sfig{sfig:justdd}c,d). These intervals use 4,000 resamples of the nine data qubits, so they describe variation
across qubits rather than shot-count uncertainty alone.

Additional state and fitted-channel comparisons are shown in \sfig{sfig:justdd}b,e,f. The superlinear increase
in flip probability without an echo is consistent with a coherent phase contribution: a phase that accumulates
across repeated windows can initially produce an approximately quadratic increase in flip probability. The
reduction with the echo is consistent with suppression of this contribution. The measured state-change
probabilities support a trade-off between phase protection and additional bit flips.

\section{Decoders and noise models}\label{sn:app_decoding}

\subsection{Decoder and noise-model comparisons}\label{sn:decoders}

We compare MWPM\cite{dennis2002topological,higgott2025sparse}, correlated
matching\cite{fowler2013optimal,higgott2025sparse}, belief propagation followed by matching\cite{higgott2023improved},
Tesseract\cite{beni2025tesseract} and exact maximum-likelihood decoding\cite{bravyi2014efficient,piveteau2024tensor}
on the selected full-patch placements with protected distances 3--6 and other
distance 3. The noise models are the calibration-derived model, a Spitz pair fit \cite{spitz2018adaptive}, and a joint
moment fit. The Spitz pair fit estimates $p_{ij}$ for pairs connected by the calibration-derived matching graph and
assigns boundary probabilities to account for remaining detector firing rates. The joint moment fit retains the
calibration-derived model's mechanisms and fits their probabilities to detector moments up to fourth order, with its
prior width chosen on a validation split. \rev{Within each noise model, every decoder is evaluated on the same shots. \methref{sec:m_noise} describes the calibration-derived model.}

\begin{figure*}[!tb]
\centering
\includegraphics[width=\linewidth]{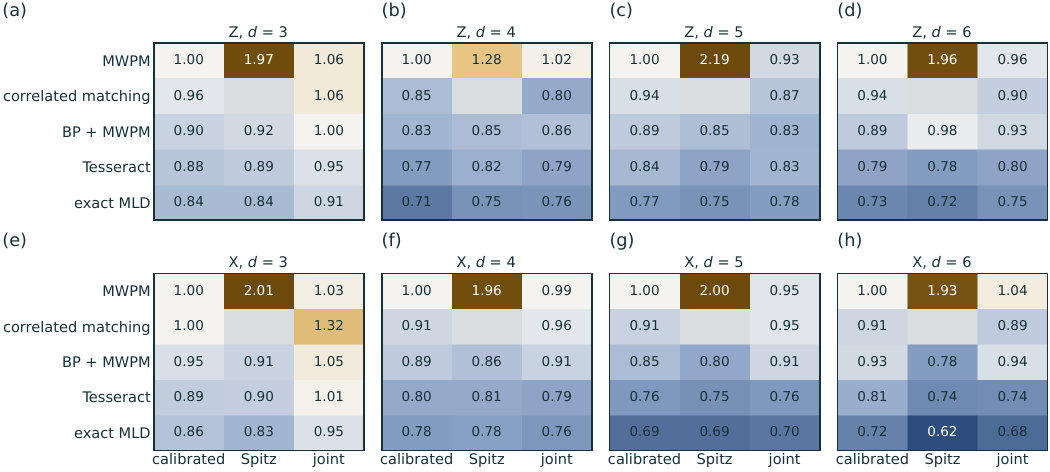}
\caption{\textbf{Logical failures by decoder and model.}
The effect of changing the noise model depends on the decoder.
\pan{a}--\pan{d} Z memory and \pan{e}--\pan{h} X memory, for protected distances 3--6 at other distance 3 after 13 rounds.
Cells give failures relative to calibration-derived MWPM on the same shots: all \num{20000} shots for the
calibration-derived model and the \num{10000} held-out shots for the two fitted models.
\rev{Rows identify decoders and columns noise models, and grey cells are undefined.}}
\label{sfig:decoders}
\end{figure*}

The comparison in \sfig{sfig:decoders}a--h reports logical failures after 13 rounds, normalized to
calibration-derived MWPM on the same shots. \rev{Entries for the calibration-derived model use all 20,000 shots, whereas the two fitted models are trained on half of the shots and evaluated on the other 10,000.}

On all 20,000 shots, calibration-derived exact decoding gives 0.69--0.86 of the MWPM failures across protected
distances 3--6, and Tesseract 0.76--0.89. At protected distance 6, exact decoding gives 498 failures against
679 for MWPM in Z memory and 462 against 641 in X memory.

The effect of fitting a noise model depends on the decoder. On the 10,000 held-out distance-six records, MWPM
with the Spitz fit gives 670 versus 342 failures in Z memory and 648 versus 336 in X memory compared with
calibration-derived MWPM. The same fitted model performs substantially better with decoders that retain more of its
error structure: exact decoding with the Spitz fit gives 245 failures in Z memory and 209 in X memory. \rev{The joint-fit results also depend on the decoder and patch, and the figure shows the individual comparisons.}

\subsection{Pair-fitted graph and boundary probabilities}\label{sn:spitz}

\begin{figure}[!tb]
\centering
\includegraphics[width=\linewidth]{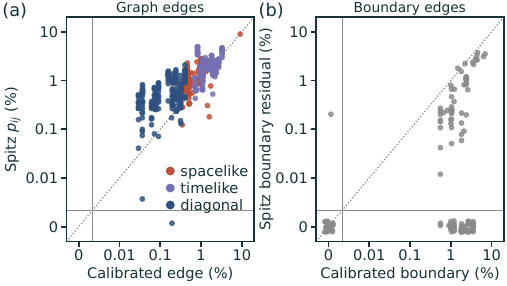}
\caption{\textbf{Graph and boundary probabilities.}
The pair fit increases many graph-edge probabilities while leaving zero boundary residuals on many detectors.
\pan{a} Spitz pair estimates versus calibration-derived graph-edge probabilities, coloured by error class.
\pan{b} Spitz boundary residuals versus calibration-derived boundary probabilities.
\rev{Dotted lines mark equality. Zeros occupy the labelled floor and are offset slightly for visibility.}
The comparison uses the model-fitting half of the shots of a 13-round X-memory circuit with protected distance 4 and other distance 3.}
\label{sfig:spitz}
\end{figure}

The diagnostic in \sfig{sfig:spitz} uses the model-fitting half of the shots of a 13-round X-memory circuit with
protected distance 4 and other distance 3. For the calibration-derived matching-graph edges, the median ratios of Spitz
pair estimates to calibration-derived probabilities are 1.77 for space-like edges, 1.18 for timelike edges and 3.86 for
diagonal edges (\sfig{sfig:spitz}a).

The pair fit also changes the boundary probabilities. It assigns zero boundary residual to 63 of the 116
detectors that have a boundary edge in the calibration-derived model (\sfig{sfig:spitz}b). These are properties of the
fitted model, rather than direct measurements of a physical boundary process.

Reducing boundary probabilities or omitting the corresponding edges changes whether MWPM matches a detection
event to the boundary or to another event. This provides a plausible explanation for its poorer performance with
the pair fit. The comparison illustrates that fitting selected detector statistics more closely need not improve
every decoder.

\end{document}